\documentstyle[aps,graphicx]{revtex}

\begin{document}

\title{Mesonic fluctuations in a nonlocal Nambu--Jona-Lasinio model}
\author{Robert S. Plant\footnote{Current address: Department of Meteorology, 
University of Reading, Reading, RG6 6BB, U.K.} and Michael C. Birse}
\address{Theoretical Physics Group, Department of Physics and Astronomy,\\
University of Manchester, Manchester, M13 9PL, U.K.}
\maketitle

\begin{abstract}
The effects of meson fluctuations are studied in a nonlocal generalization of 
the Nambu--Jona-Lasinio model, by including terms of next-to-leading order 
(NLO) in $1/N_c$. In the model with only scalar and pseudoscalar interactions 
NLO contributions to the quark condensate are found to be very small. This is
a result of cancellation between virtual mesons and Fock terms, which occurs 
for the parameter sets of most interest. In the quark self-energy, similar 
cancellations arise in the tadpole diagrams, although not in other NLO pieces
which contribute at the $\sim 25\%$ level. The effects on pion properties are 
also found to be small. NLO contributions from real $\pi\pi$ intermediate
states increase the sigma meson mass by $\sim 30\%$.
 In an extended 
model with vector and axial interactions, there are indications that NLO 
effects could be larger.
\end{abstract}


\section{Introduction}

Until recently most calculations performed with the Nambu--Jona-Lasinio (NJL) 
model\cite{NJL,NJLrevs} have been restricted to tree-level in the mesons or, 
in other words, to leading order (LO) in the $1/N_c$ expansion. Although the
possible importance of quantum fluctuations of the mesons has been generally
recognized (see, for example, Refs.\cite{QK,nlo_rhodecay,nlo}), a satisfactory
treatment of such effects has required the development of schemes that are
faithful to the chiral invariance of the model. This is crucial, given that
the most important motivation for studying the model is its description of the
dynamical symmetry breakdown. Schemes have been introduced on the basis of an
effective action method\cite{NBCRG}, an appropriate selection of Feynman 
diagrams\cite{Bowler,DSTL} and a bosonized approximation\cite{P97}.

Calculations at next-to-leading order (NLO) in $1/N_c$ are certainly more 
involved than at LO, both analytically and numerically. However, since 
$1/N_c$ is such a modest expansion parameter, it has rightly been seen as 
important to estimate the size of some NLO effects. Even if the calculations 
were of interest for no other reason, a check on this perturbative approach 
would be very valuable. Encouragingly, there seems to be a general consensus 
emerging from recent studies that NLO corrections are indeed relatively small
for standard choices of the model 
parameters\cite{NBCRG,DSTL,nlo_chi,qmatter,oertel,nlo_rho,ripka,oertel_thesis},
at least 
for properties of the ground state and of the pion. A contrary view, however,
has been expressed by Kleinert and Van den Bossche\cite{kle}.

The most compelling reason for a NLO analysis, however, is that the LO 
treatment of an NJL-type model neglects physical processes that are expected 
to be qualitatively important. For instance, several of the particles 
described by such models ($\sigma$, $\rho$, $a_1$) are broad resonances, yet 
the model meson propagators at LO are purely real. At NLO the particle widths
are incorporated in a natural way, by including diagrams with two-meson 
intermediate states in the Bethe--Salpeter equation (BSE). Although these 
widths can be estimated from the LO decay amplitudes, the effects on the 
position of a resonance require a NLO treatment. Such effects may well prove 
important in model descriptions of, for example, the $\rho$ 
meson\cite{nlo_rhodecay,nlo_rho}.

A difficulty with the model at NLO concerns the specification of model 
parameters associated with regularization of the loop integrals. At LO the 
model contains three parameters: the current quark mass, a coupling constant 
and a cutoff parameter that regularizes quark loops. Conventionally two of 
these parameters are fitted to the pion mass and decay constant, while the
third is at least constrained by the value of the quark condensate. However,
the regularization of the model must be further specified at NLO since meson
and quark loop integrals can be regulated quite independently without 
affecting chiral symmetry 
relations\cite{NBCRG,DSTL,P97,oertel,nlo_rho,ripka,newpreprint}. The 
parameter space can 
be limited to some extent because the NLO corrections produce instabilities 
for large values of the meson cutoff\cite{nlo_rho,newpreprint}. Nonetheless 
the freedom 
to introduce another cutoff parameter produces significant uncertainty. 
Oertel {\em et al.}\cite{nlo_rho} have determined the meson cutoff by 
studying the pion electromagnetic form factor and find values well away from 
the region of instabilities. This is a promising result, but to put it on a 
firmer basis one would have to relax assumptions that were made in 
Ref.\cite{nlo_rho} for numerical reasons\footnote{The $\rho$ meson was
included in the model but vector and axial-vector degrees of freedom were not
included as intermediate states in the NLO diagrams. In addition, $\pi a_1$
mixing was neglected, even though this is a LO effect.}. An attractive
alternative is to perform the regularization at the level of the model action
rather than in the loop integrals. In this approach the regularization is
specified from the outset and does not have to be imposed at each order of the
expansion, a point which has been stressed by Ripka\cite{ripka}.

In this paper, we make a NLO analysis of a model that was originally proposed 
in Ref.\cite{BB95}. It has subsequently been studied at LO in the 
meson\cite{nonlocal,Broniowski}, baryon (soliton)\cite{nlsoliton} and quark 
matter\cite{qmatter} sectors. The two-body interaction vertex is nonlocal and
separable, a Gaussian form factor being associated with each quark field. Such
a model successfully eliminates several of the traditional problems of the NJL
model whilst nevertheless retaining much of the simplicity that is its chief
merit. The separable nature of the interaction is motivated in part by
instanton liquid  studies\cite{DP,Musakhanov}. The interaction form factors
ensure the convergence of all loop integrals and so the NLO corrections are
unambiguous. In addition, a practical advantage is conferred by the shape of
the form factor, which allows complicated NLO diagrams to be evaluated
efficiently with Gaussian numerical techniques. Such properties make this
model a particularly convenient one with which to examine NLO effects. NLO 
quantum fluctuations of the quark condensate in a variant of this model were 
studied in Ref.\cite{ripka}. Here we investigate the NLO treatment of mesons.

In our approach we keep the full momentum dependence of all quark loop
diagrams, and avoid expanding them in powers of meson momenta. Such an 
expansion may be useful in the determination of pion properties, but it is 
not valid for heavy mesons, such as the scalar, isoscalar sigma.

\section{The nonlocal model}\label{lo_review}

In this section we define the model used and recall some important aspects of
its behaviour at LO. This is entirely standard\cite{NJLrevs,nonlocal} and is 
included largely in order to establish our notation. The action of a quark 
model with a four-point interaction may be written
\begin{displaymath}
S= \int\! d^4\!x\, \overline{\psi}(x) (i {\rlap/\partial} - m_c) \psi (x) 
+ \sum_i\int \prod_n d^4\! x_n\, H_{i} (x_1,x_2,x_3,x_4)\,
\end{displaymath}
\begin{equation}
\times \, \overline{\psi}(x_1) \Gamma^\alpha_i \psi (x_3) \overline{\psi}
(x_2) \Gamma_{i\alpha} \psi (x_4) , \label{4q action}
\end{equation}
where $\Gamma^\alpha_i$ denotes Dirac, colour and isospin matrices. Chiral
symmetry constrains certain of the possible Dirac and isospin structures to
appear in particular combinations. The admissible interaction terms are 
\begin{eqnarray}
H_1(1 \otimes 1 + i\gamma_5 \tau^a \otimes i\gamma_5 \tau^a), &&\qquad 
H_2(\gamma^\mu \tau^a \otimes \gamma_\mu \tau^a + \gamma^\mu \gamma_5 \tau^a 
\otimes \gamma_\mu \gamma_5 \tau^a), \nonumber\\
H_3(\gamma^\mu \otimes \gamma_\mu), &&\qquad H_4(\gamma^\mu \gamma_5 \otimes 
\gamma_\mu \gamma_5), \nonumber\\
H_5(\tau^a \otimes \tau^a + i\gamma_5 \otimes i\gamma_5), &&\qquad 
H_6(\sigma_{\mu\nu} \otimes \sigma^{\mu\nu} - \sigma_{\mu\nu} \tau^a \otimes 
\sigma^{\mu\nu} \tau^a) . 
\end{eqnarray}
We shall be primarily concerned with the simplest version of the model,
including the pions, and their chiral partner, through the $H_1$ interaction.
In the usual NJL model the $H_i$ are simply constants. However, the use of a 
nonlocal interaction is appealing\cite{BB95,nonlocal,nlsoliton}, not least 
because it enables one to avoid some of the ambiguities associated with
regularization. Moreover, a nonlocal interaction of separable form is 
suggested by studies of the instanton-liquid picture of the QCD 
vacuum\cite{DP,Musakhanov}. In this case, the momentum space form of the 
$H_i$ is
\begin{equation}
H_i (p_1 ,p_2 ,p_3 ,p_4 )= {\textstyle\frac{1}{2}} (2\pi)^4 G_i f(p_1) f(p_2) 
f(p_3) f(p_4) \delta (p_1 + p_2 - p_3 - p_4) , \label{separable}
\end{equation}
with the $G_i$ being constants which, for the purposes of counting, we take 
to be ${\cal O} (1/N_c)$. Although the instanton-liquid model 
\cite{DP,Musakhanov} 
predicts a particular form for the form-factor function $f(p)$, other
authors\cite{BB95,nonlocal,nlsoliton} have considered a function which is a 
Gaussian in Euclidean space, 
\begin{equation}
f(p_E)=\exp(-p_E^2 / \Lambda^2 ). \label{form factor}
\end{equation}
The Gaussian shape produces a good phenomenology of mesons and nucleons in the
LO approximation\cite{nonlocal,nlsoliton} and is used for the numerical 
results in this paper.

At leading order in the $1/N_c$ expansion the quark propagator is dressed by
a single quark-loop self-energy diagram. This results in a momentum-dependent
quark ``mass", $m(p)$:
\begin{equation}
S^{-1} (p)={\rlap/p} - m(p) ; \label{q prop}
\end{equation}                 
\begin{equation}
m(p)=m_c + \Bigl(m(0) - m_c\Bigr)f^2(p) ,
\end{equation}
where the constant $m(0)$ is obtained from the Schwinger--Dyson equation 
(SDE)\cite{BB95}. If $m(0)$ is sufficiently large, the poles in the quark 
propagator are shifted into the complex $p^2$ 
plane\cite{ripka,BB95,nonlocal}. A feature of models with nonlocal 
interactions is the appearance of additional, unphysical poles in the quark 
propagator. For the parameter sets of most interest these are located well 
away from the real axis\cite{nonlocal}, but instabilities of the model ground
state can appear for small values of the cutoff\cite{ripka}.

Mesonic bound states can be found from the poles in the $\overline{q}q$
scattering matrix, $T$. For a separable interaction, it is convenient to
factor out interaction form factors of the external quark momenta and to 
define
\begin{equation}
T(p_1,p_2,p_3,p_4) = \delta(p_1 + p_2 - p_3 - p_4) \prod_n f(p_n)
\sum_{i,j}\Gamma_i\hat T_{ij}(q)\overline{\Gamma}_j, \label{BSE hat}
\end{equation}
where $q=p_1- p_3=p_4 - p_2$ denotes the total momentum of the $\overline{q}q$
pair. The notation $\overline{\Gamma}_j$ is needed in the longitudinal
components of the vector and axial channels where, for example, 
$\Gamma_V=-\overline\Gamma_V=i{\rlap/q}$. In all other cases $\Gamma_j$ and
$\overline{\Gamma}_j$ are the same (see Eq.~(23) of Ref.\cite{nonlocal}). The
Bethe--Salpeter equation (BSE) at LO is given by\cite{BB95,nonlocal}
\begin{equation}
\hat T(q)=G + GJ(q)\hat T(q) , \label{BSE matrix}
\end{equation}
where $G$ is simply a matrix of the coupling constants from the action and
$J(q)$ is the matrix of polarization loop integrals
\begin{equation}
J_{ij}(q)=i \hbox{Tr} \int \frac{d^4 p}{(2\pi)^4} f^2(p_+) f^2(p_-) \Gamma_j 
S(p_-) \overline{\Gamma}_i S(p_+),  \label{J}
\end{equation}
where we have introduced $p_\pm=p\pm {1\over 2}q$. The symbol `Tr' is used to 
denote a trace over flavour, colour and Dirac indices.

Close to the pole corresponding to a particular particle, the amplitude in the
relevant channel may be written
\begin{equation}
\frac{\overline{V}(q) \otimes V(q)}{m^2 - q^2}, \label{vertex fn}
\end{equation}
where $V(q)$ denotes the particle vertex function. In the simplest form of the
model, with only the $G_1$ coupling included, these functions are:
\begin{equation}
V_\pi (q)=g_{\pi qq} i\gamma_5 \tau^a, \qquad V_\sigma (q)= g_{\sigma qq}, 
\end{equation}
where:
\begin{equation}
{1\over g_{iqq}^2}=\left. \frac{dJ_{ii}}{dq^2} \right|_{q^2=m^2}.
\label{meson couplings}
\end{equation}
The vertex functions appear in calculations of mesonic properties, such as the
pion decay constant, $f_\pi$. 

In order to obtain symmetry currents with the same divergences as the 
corresponding local currents in QCD, and hence to respect the corresponding 
Ward identities, one has to include additional nonlocal terms in the
currents\cite{BB95,nonlocal,nlsoliton,currents}. These arise as a consequence 
of the nonlocality of the action. Details of the nonlocal terms required in 
both vector and axial currents can be found in Refs.\cite{BB95,nonlocal}. 
Although the method developed in those papers is not unique and there is an 
ambiguity in defining the transverse parts of the currents, this does not 
affect the longitudinal components.

\begin{figure}
\begin{center}
\includegraphics[bb=188 645 478 692]{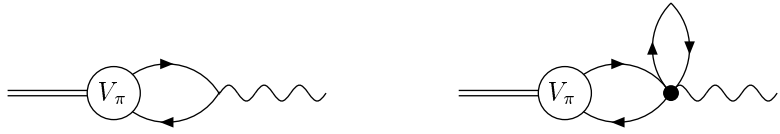}
\end{center}
\caption{The LO couplings of a pion to the axial current (represented by the 
wavy line).}
\label{LO_pi-axial}
\end{figure}

As an example, consider the pion decay constant. The diagrams in 
Fig.~\ref{LO_pi-axial} represent the contributions from the local and nonlocal
pieces of the axial current. The longitudinal part of the current, $q^\mu
A_\mu^a$, contains the following term
\begin{eqnarray}
\frac{i}{2(2\pi)^{12}} G_1 \int && \prod_n d^4 p_n\, \overline{\psi} (p_1) i 
\gamma_5 \tau^a \psi (p_3) \overline{\psi} (p_2) \psi (p_4) ~\, \delta(p_1 + 
p_2 + q - p_3 - p_4)\nonumber\\
&&\!\!\!\times \left[ f(p_1)f(p_2)f(p_3)f(p_4-q) + f(p_1)f(p_2)f(p_3-q)f(p_4) 
\right.\nonumber\\
&&\!\!\!\left. -f(p_1)f(p_2+q)f(p_3)f(p_4) - f(p_1+q)f(p_2)f(p_3)f(p_4) 
\right] . \label{lo_current_example}
\end{eqnarray}
This term contributes to the pion decay constant since the operator
$\overline{\psi} (p_2) \psi (p_4)$ has a nonzero vacuum expectation value
(represented by the bubble loop in the second diagram of 
Fig.~\ref{LO_pi-axial}). The rest of the operator, $\overline{\psi} (p_1) i 
\gamma_5 \tau^a \psi (p_3)$, contributes to the pion-to-vacuum matrix element
of the current. Such terms are needed in order to satisfy the 
Gell-Mann--Oakes--Renner (GMOR) relation\cite{GMOR}, as discussed in 
Ref.~\cite{BB95}, and they can make significant numerical contributions to 
various observables\cite{nonlocal}.

\section{NLO contributions}\label{NLO diagms}
\subsection{Symmetry currents}\label{exchange_curr}

At leading order in the $1/N_c$ expansion the fields at $x_1$ and $x_3$ in 
the interaction term of the action (\ref{4q action}) are always connected to 
the same quark loop, and similarly for the fields at $x_2$ and $x_4$. Beyond 
this order, there are also exchange or ``Fock" terms where the fields at $x_1$
and $x_4$ are connected to the same loop. These can be constructed by first 
performing a Fierz transformation on the action and then using the
transformed action in just the same way that one uses the original action at 
LO. The Fierz transformed action of this model contains the following 
colour-singlet terms:
\begin{eqnarray}
&&(4N_c)^{-1} (G_1 - 2G_3 + 2G_4 - G_5 + 12G_6) \, 
(1 \otimes 1 + i \gamma_5 \tau^a \otimes i \gamma_5 \tau^a), \nonumber\\
&&(4N_c)^{-1} (-2G_2 + G_3 + G_4) \, 
(\gamma_\mu \tau^a \otimes \gamma^\mu \tau^a 
+ \gamma_\mu \gamma_5 \tau^a \otimes \gamma^\mu \gamma_5 \tau^a), \nonumber\\
&&(4N_c)^{-1} (-2G_1 + 6G_2 + G_3 + G_4 - 2G_5) \, 
(\gamma_\mu \otimes \gamma^\mu), \nonumber\\
&&(4N_c)^{-1} (2G_1 + 6G_2 + G_3 + G_4 + 2G_5) \, 
(\gamma_\mu \gamma_5 \otimes \gamma^\mu \gamma_5), \nonumber\\
&&(4N_c)^{-1} (-G_1 - 2G_3 + 2G_4 + G_5 - 12G_6) \, 
(\tau^a \otimes \tau^a + i \gamma_5 \otimes i \gamma_5), \nonumber\\
&&(8N_c)^{-1} (G_1 - G_5 - 4G_6) \, (\sigma_{\mu\nu} \otimes \sigma^{\mu\nu}
- \sigma_{\mu\nu} \tau^a \otimes \sigma^{\mu\nu} \tau^a) . \label{fierz}
\end{eqnarray} 
(There are also colour-octet terms which we do not use here.)

The nonlocal terms in the symmetry currents of the model also involve four
quark fields and so will also be subject to exchange effects at NLO. These 
Fock pieces in the currents will introduce further ambiguity through the 
definition of their transverse parts. One way to construct the NLO current 
terms is to take the Fierz transformed action and to apply the same method as
was used in Refs.\cite{BB95,nonlocal} to form the nonlocal currents from the
original model action. This method leads to nonlocal NLO current terms of the
same structures as those presented in Ref.\cite{nonlocal}, with the 
appropriate coupling constants in the original terms replaced by the 
corresponding combinations of couplings from the Fierz transformed action. 
For example, as well as a term $G_1 (i \gamma_5 \tau^a \otimes 1)$ in the 
nonlocal axial current constructed from the original action (see 
Eq.~(\ref{lo_current_example})), there is a Fock term of the same structure 
where $G_1$ is replaced by $(4N_c)^{-1} (G_1 - 2G_3 + 2G_4 - G_5 + 12G_6)$.

An alternative approach to constructing the Fock terms of the model's 
currents is simply to Fierz transform the LO currents of Ref.\cite{nonlocal}.
The differences between this procedure and the one outlined above (where the
Noether-like method for deriving a symmetry current is applied after the Fierz
transformation) lie in the combinations of form factors which are present in
the currents. However, these affect only the transverse parts of the currents;
both definitions satisfy the appropriate Ward identities. The equivalence of
the longitudinal components has been checked explicitly\cite{thesis}. In 
practice, the currents obtained from the Fierz-transformed action are 
somewhat easier to work with, since they retain the same structures as the
nonlocal terms in the LO currents, with appropriate substitutions for the
overall coefficients.
  
\subsection{Quark propagator}\label{SDE_NLO}

We are now in a position to consider the quark and meson propagators at NLO. 
Like Oertel {\em et al.}\cite{oertel,nlo_rho,oertel_thesis} and 
Ripka\cite{ripka} we use a strict NLO scheme in which all propagators and 
matrix elements are expanded to NLO in $1/N_c$. This scheme has also been 
presented by Dmitra\u{s}inovi\'{c} {\em et al.}\cite{DSTL}, who themselves 
preferred to adopt an alternative scheme where certain NLO terms in the 
quark self-energy are treated self-consistently in the quark Schwinger--Dyson 
equation. Their scheme resums a subset of terms to all orders in $1/N_c$.
Both schemes satisfy chiral low-energy theorems\cite{DSTL,nlo_rho}. 
The Feynman diagrams required by the strict NLO scheme in the nonlocal NJL 
model are briefly presented below. 

At NLO the quark Schwinger--Dyson equation is coupled to the meson 
Bethe--Salpeter equation and takes the form\cite{Bowler}
\begin{eqnarray}
S_F^{-1} (p)&=& {\rlap/p} - m_c - iG_1 f^2(p) {\mathrm{Tr}} \int 
\frac{d^4k}{(2\pi)^4} S_F(k) f^2(k)\nonumber\\
&&\qquad\qquad+ i f^2(p) \sum_{i,j} \int \frac{d^4k}{(2\pi)^4} 
\hat T_{ij}(k) \Gamma_i S_F (p-k) \overline{\Gamma}_j f^2(p-k) , 
\label{SDE NLO}
\end{eqnarray}
where $\hat T_{ij}(k)$ denotes the element of the $\overline{q}q$ scattering 
matrix which describes scattering from the state with matrix structure
$\Gamma_j$ to the state with structure $\Gamma_i$. This matrix is diagonal 
apart from the mixing between the pseudoscalar and axial channels which is 
present in the extended model with vector and axial interactions. It is to be
understood as the LO scattering matrix, which is of order $N_c^{-1}$ (see 
Eq.~(\ref{BSE matrix})). 

Although the SDE as written above contains all of the required terms at LO 
and NLO, it also includes some unwanted higher order terms. The NLO terms can
be obtained by replacing the quark propagator in the final integral by its LO
form, and by expanding the propagator in the quark ``bubble" integral to NLO:
\begin{equation}
S_F (p) \simeq S(p) + S(p)\Sigma^{(1)} (p)S(p) , \label{prop decomposition}
\end{equation}
where $\Sigma^{(1)}$ denotes the NLO contribution to the quark self-energy. 
This leads to an equation for $\Sigma^{(1)}$ of the form
\begin{eqnarray}
\Sigma^{(1)} (p) &=& iG_1 f^2(p) {\mathrm{Tr}} \int \frac{d^4k}{(2\pi)^4} S(k)
\Sigma^{(1)}(k)S(k) f^2(k)\nonumber\\
&& - i f^2(p) \sum_{i,j} \int \frac{d^4k}{(2\pi)^4} \hat T_{ij}(k) 
\Gamma_i S (p-k) \overline{\Gamma}_j f^2(p-k) . \label{qprop NLO}
\end{eqnarray}
The first term comes from an insertion of the NLO self-energy into the quark
bubble diagram. The second, involving the scattering matrix, includes both
Fock terms and dressing of the quark by virtual mesons.

The separable nature of the interaction means that this equation can be
solved straightforwardly to get
\begin{equation}
\Sigma^{(1)} (p) = f^2(p)C - i f^2(p) \sum_{i,j} \int \frac{d^4k}{(2\pi)^4} 
\hat T_{ij}(k) \Gamma_i S (p-k) \overline{\Gamma}_j f^2(p-k) , 
\label{qprop NLO decomp}
\end{equation}
where
\begin{eqnarray}
C &=& \frac{G_1}{1 - G_1 J_{\sigma\sigma} (0)} \sum_{i,j} {\mathrm{Tr}} \int 
\frac{d^4k}{(2\pi)^4} \int \frac{d^4\ell}{(2\pi)^4} 
\hat T_{ij}(k-\ell)\nonumber\\
&&\qquad\qquad\qquad\times S(k) \Gamma_i S(\ell) \overline{\Gamma}_j S(k) 
f^4(k) f^2(\ell) . \label{c defn}
\end{eqnarray}
The two terms correspond to the tadpole and meson cloud diagrams shown in 
Fig.~\ref{NLO_q-self}. The tadpole diagram is responsible for the contribution
$f^2(p)C$. It includes the exchange of a zero-momentum sigma meson between
the quark and a virtual meson. Since $C$ is a momentum-independent constant,
this contribution has the same form as the LO quark self-energy. The other
diagram describes the emission and subsequent reabsorption of a virtual meson.
It has a nontrivial dependence on the momentum of the quark and generates a
wave function renormalization as well as a scalar term. Note that although we
shall often refer to the upper double line in such diagrams as a meson
propagator, it is really a $\overline{q}q$ scattering amplitude and so the
diagrams contain Fock terms as well as virtual meson contributions.

\subsection{Meson propagators}\label{BSE_NLO}
         
At LO, the mesonic bound states are constructed from the ladder 
Bethe--Salpeter equation (Eq.~(\ref{BSE matrix})). The NLO extension of the 
BSE can be expressed in terms of corrections to the basic two-quark 
loop\footnote{We refer to a loop integral containing $n$ LO quark propagators
as an $n$-quark loop.} of Eq.~(\ref{J}). If the quantities $J_{ij}$ are 
redefined to include these NLO contributions, the scattering matrix can still
be written in the form of Eqs.~(\ref{BSE hat}) and (\ref{BSE matrix}). 

An obvious set of NLO terms in a polarization loop integral $J_{ij}(q)$
consists of insertions of the NLO quark self-energies of Fig.~\ref{NLO_q-self}
on either the quark or antiquark line. These contribute
\begin{eqnarray}
&&i {\mathrm{Tr}} \int \frac{d^4p}{(2\pi)^4} \Gamma_j S(p_-) \Sigma^{(1)} 
(p_-)S(p_-) \overline{\Gamma}_i S(p_+) f^2(p_-) f^2(p_+)\nonumber\\
&+&i {\mathrm{Tr}} \int \frac{d^4p}{(2\pi)^4} \Gamma_j S(p_-) 
\overline{\Gamma}_i 
S(p_+) \Sigma^{(1)} (p_+) S(p_+) f^2(p_-) f^2(p_+) . \label{BSE NLO qprop}
\end{eqnarray}

\begin{figure}
\begin{center}
\includegraphics[bb=194 607 473 689]{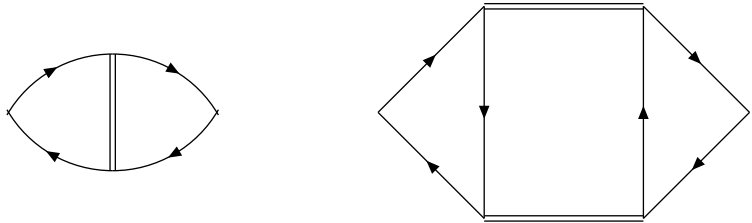}
\end{center}
\caption{Meson exchange and two-meson diagrams in the BSE at NLO. There is 
also a second two-meson diagram (not shown) in which the direction of the 
arrows in one of the quark loops is reversed.}
\label{NLO_BSE}
\end{figure}

A second kind of NLO contribution arises from the exchange of a $t$-channel
virtual meson between the quark and the antiquark. The corresponding diagram 
is shown in Fig.~\ref{NLO_BSE} and makes the following contribution to 
$J_{ij}(q)$:
\begin{eqnarray}
&&\sum_{l,m} {\mathrm{Tr}} \int \frac{d^4p}{(2\pi)^4} \int 
\frac{d^4k}{(2\pi)^4} \hat T_{lm}(p-k) \Gamma_j S (k_-) \Gamma_l S (p_-) 
\overline{\Gamma}_i S (p_+) \overline{\Gamma}_m S (k_+)\nonumber\\
&&\qquad\qquad\qquad\times f^2(k_-) f^2(k_+) f^2(p_-) f^2(p_+) . 
\label{BSE NLO 1mes}
\end{eqnarray}

Finally, there are NLO contributions that involve intermediate two-meson
states, represented by the second diagram in Fig.~\ref{NLO_BSE}. These allow 
for the instability of a meson against two-body decays and so can introduce an
imaginary part in the corresponding propagator above the threshold energy for 
the two-particle final states. They are constructed by connecting two LO
three-meson vertices with two meson propagators. The resulting imaginary part
of a meson mass can be expressed in terms of the square of the LO decay 
amplitude. These contributions also generate a shift in the real part of the 
mass which involves a new type of loop integral.

In writing explicit expressions for these contributions, it is convenient to 
introduce functions $L_{ij,k}$ and $\overline{L}_{i,jk}$ to describe the LO 
three-point vertices. The function $L_{ij,k}$ describes the process $k\to ij$ 
and is defined as
\begin{eqnarray}
L_{ij,k}(q_1,q_2,q)&=& i\,{\mathrm{Tr}}\int \frac{d^4p}{(2\pi)^4} \Gamma_k 
S(p_-) \overline{\Gamma}_j S( p - {\textstyle\frac{1}{2}} (q_1 - q_2)) 
\overline{\Gamma}_i S(p_+)\nonumber\\
&&\qquad\qquad \times f^2(p_-) f^2(p_+) f^2 ( p - {\textstyle\frac{1}{2}} 
(q_1 - q_2) )\nonumber\\
&&+ i\, {\mathrm{Tr}} \int \frac{d^4p}{(2\pi)^4} \Gamma_k S(p_-) 
\overline{\Gamma}_i 
S ( p + {\textstyle\frac{1}{2}} (q_1 - q_2)) \overline{\Gamma}_j S(p_+)
\nonumber\\
&&\qquad\qquad\times f^2(p_-) f^2(p_+) f^2 ( p + {\textstyle\frac{1}{2}} 
(q_1 - q_2)) ,
\end{eqnarray}
while $\overline{L}_{i,jk}$ describes the process $jk \to i$,
\begin{eqnarray}
\overline{L}_{i,jk}(q,q_1,q_2)&=& i\, {\mathrm{Tr}} \int 
\frac{d^4p}{(2\pi)^4} \overline{\Gamma}_i S(p_+) \Gamma_j 
S ( p - {\textstyle\frac{1}{2}} (q_1 - q_2) ) \Gamma_k S(p_-)\nonumber\\
&&\qquad\qquad\times f^2(p_-) f^2(p_+) f^2 ( p - {\textstyle\frac{1}{2}} 
(q_1 - q_2))\nonumber\\
&&+ i\, {\mathrm{Tr}} \int \frac{d^4p}{(2\pi)^4} \overline{\Gamma}_i S(p_+) 
\Gamma_k 
S ( p + {\textstyle\frac{1}{2}} (q_1 - q_2) ) \Gamma_j S(p_-)\nonumber\\
&&\qquad\qquad\times f^2(p_-) f^2(p_+) f^2 ( p + {\textstyle\frac{1}{2}} 
(q_1 - q_2) ) .
\end{eqnarray}
The contribution to $J_{ij}(q)$ from the two-meson diagrams can then be
written in the form
\begin{equation}
-\,\frac{i}{2} \sum_{k,l,m,n} \int \frac{d^4p}{(2\pi)^4} 
\overline{L}_{i,km}(q,p_+,-p_-) \hat T_{kl}(p_+) \hat T_{mn}(-p_-) 
L_{ln,j}(p_+,-p_-,q) . \label{BSE NLO 2mes}
\end{equation}                          

Once the full NLO matrix $J^{(1)}$ has been constructed, we can use it to
find the corrections to meson masses and meson-quark couplings. As an example,
consider the pion in the version of the model without vector and axial
couplings. The pion pole is located at
\begin{equation}
1 - G_1 \left[J^{(0)}_{\pi\pi} (q^2) + J^{(1)}_{\pi\pi} (q^2)\right] =0, 
\label{pion shell}
\end{equation}
where superscripts $(0)$ and $(1)$ denote LO and NLO terms. The coupling to 
quarks (Eq.~(\ref{meson couplings})) is given by 
\begin{equation}
\left(g^{(0)}_{\pi qq}+g^{(1)}_{\pi qq}\right)^{-2} = \left. 
\frac{d (J^{(0)}_{\pi\pi} + J^{(1)}_{\pi\pi})}{dq^2} \right|_{q^2=m_\pi^2}.
\end{equation}
From these we get the squared pion mass to NLO,
\begin{equation}
m_\pi^2=m_\pi^{(0)\,2}\left[1+g_{\pi qq}^{(0)\,2}\, 
\frac{d J^{(1)}_{\pi\pi}}{d q^2}\right]_{q^2=m_\pi^{(0)\,2}}
- g_{\pi qq}^{(0)\,2} J^{(1)}_{\pi\pi} (m_\pi^{(0)\,2}),
\end{equation}
and the pion-quark coupling,
\begin{equation}
g_{\pi qq}^{(1)}= -\,\frac{g_{\pi qq}^{(0)\,3}}{2} \left[
2m_\pi^{(0)} m_\pi^{(1)}\frac{d^2 J^{(0)}_{\pi\pi}}{d(q^2)^2}
+\frac{d J^{(1)}_{\pi\pi}}{dq^2} \right]_{q^2=m_\pi^{(0)\,2}} . 
\label{g_Npiqq}
\end{equation}

\subsection{Meson coupling to a current}\label{fpi_NLO}

In this section, we consider a NLO determination of the coupling between a
meson and an external current, taking the pion decay constant as an example.
The situation is more complicated in a nonlocal than a local
model\cite{NBCRG,DSTL} because there are two kinds of contribution to be
considered, arising from the local and nonlocal parts of the symmetry current
(see Sec.~\ref{lo_review}). 

The LO diagrams are shown in Fig.~\ref{LO_pi-axial}. At NLO several of the
extra contributions can be identified straightforwardly, either by inserting
the NLO self-energy on any one of the quark lines in these diagrams or by
using the NLO contribution to the pion vertex function. Other contributions 
are analogous to those in the BSE at NLO. These involve either a $t$-channel 
meson exchange in the two-quark loop or intermediate two-meson states. They 
are shown in Fig.~\ref{NLO_pi-ax_meson1}.

\begin{figure}
\begin{center}
\includegraphics[bb=263 645 393 710]{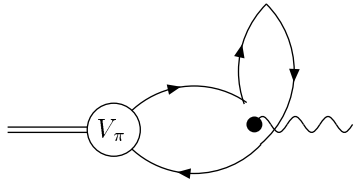}
\end{center}
\caption{The exchange contribution to the coupling of the pion to the nonlocal
axial current. Note that this Fock diagram has been distinguished from the 
similar LO diagram by separating the quark lines to indicate the flow of 
colour.}
\label{NLO_pi-ax_exch}
\end{figure}

The remaining NLO contributions arise solely from the nonlocal piece of the 
current. The first of these are Fock terms arising from the exchange of two of
the quark lines at a nonlocal current vertex, as shown in
Fig.~\ref{NLO_pi-ax_exch}. Since they contain only a single colour trace, as
indicated by separating the quark lines associated with each of the
$\overline\psi\Gamma\psi$ factors, this diagram is suppressed by one power of
$N_c$ compared with the corresponding LO terms. Although, as noted above, 
there is an ambiguity in defining the transverse components of such Fock 
terms, this does not affect the calculation of the pion decay constant. 

Finally there are two NLO contributions which can be thought of as arising
from nonlocal current couplings to virtual mesons. They are shown in 
Fig.~\ref{NLO_pi-ax_meson2}. The first (a) is somewhat similar to diagram
\ref{NLO_pi-ax_meson1}(d), except that the coupling of the nonlocal current 
to the virtual mesons involves both the quark and the antiquark instead of a 
separate quark bubble. This is analogous to the two-body diagrams that
contribute at LO to several of the electromagnetic amplitudes described in
Ref.\cite{nonlocal}. The other diagram (b) can be thought of as coupling the
current at a meson-quark vertex.

Although the NLO contributions to the pion decay constant involve rather a
large number of diagrams, in practice the situation can be significantly
simplified as a result of cancellations amongst them\cite{thesis}. A complete
description of these cancellations would be somewhat lengthy and so we offer 
a brief summary in Appendix~\ref{fpi_nlo}. We have also checked the 
consistency of our treatment by verifying that the GMOR relation\cite{GMOR} 
holds to NLO (see Appendix~\ref{gmor_nlo}).

\section{Numerical results}\label{nloresults}
\subsection{Model parameters}\label{nlofits}

The multiple integrals involved in the NLO diagrams can be rather 
time-consuming to evaluate numerically to good accuracy. We have therefore 
chosen to fit the model parameters at LO and then to calculate the NLO 
changes to observables. In fact, for the parameter sets of most interest, we 
find that the corrections are small for the observables used to fix the 
parameters. Thus, our results should not be very different from those of a 
full NLO fit.

We fit the parameters at LO as in Ref.\cite{nonlocal}. In the simpler version 
of the model with no vector and axial couplings, we fit two parameters to the 
pion mass and decay constant. This leaves one free parameter which we 
characterize in terms of $m_0(0)$, the quark self-energy at zero momentum in
the chiral limit. Details of the sets used are given\footnote{Note that the 
parameters given differ slightly from those quoted in Ref.\cite{BB95} where a
very similar fit was made at the same values of $m_0(0)$. This is simply 
because the calculations of Ref.\cite{BB95} were performed within the chiral 
expansion of the model.} in Table~\ref{tab1}. Values of some quantities 
calculated at LO with these parameter sets are also given in the table. They 
are qualitatively similar to those obtained in the extended version of the 
model\cite{nonlocal}. For the parameter sets with $m_0(0)=300$ MeV and 
larger, the quark propagator has no poles on the real axis and so single 
quarks do not appear as physical states\cite{BB95,nonlocal}. 

Although most of our work at NLO has used this simpler model, we also show a 
few results for the extended model including vector and axial couplings. For 
these we use a parameter set with $m_0(0)=300$ MeV, which is intermediate
between the two cases studied at LO in Ref.\cite{nonlocal}.

\subsection{Quark condensate}\label{num cond}

The NLO term in the quark condensate can be evaluated as the sum of two 
contributions. These correspond to the two self-energy diagrams of 
Fig.~\ref{NLO_q-self} where the external quark line is formed into a closed
loop with a local scalar insertion\cite{Bowler}. By comparing the values for 
the condensate at NLO in Table~\ref{tab2} with the LO values in 
Table~\ref{tab1}, we see that the NLO contributions are very small. This may 
seem surprising since virtual pions would be expected to make a large 
(positive) contribution to the condensate. However it should be recalled that
the meson ``propagators" in the NLO diagrams also contain the Fock terms of 
the two-body interaction. These make a negative contribution to the 
condensate which tends to cancel the contributions of virtual mesons. This 
agrees with the findings of Ref.\cite{ripka} in a similar model. In the local
version of the NJL model, the condensate is also little altered at NLO if 
proper time regularization of the quark loops is used\cite{NBCRG}, but in the
$O(4)$ scheme there can be appreciable changes at NLO. Including only the pole
pieces of the meson propagators, significant shifts in the condensate are also
found with $O(3)$ regularization\cite{schmidt}.

\subsection{Quark propagator}\label{num quark}

The quark self-energy at NLO is given by Eqs.~(\ref{qprop NLO decomp}) and 
(\ref{c defn}). Its evaluation requires integrating over the scattering
matrix $\hat T$. Evaluating the ${\overline q}q$ loop integrals $J$ for each 
point in the mesonic loop integral is very time consuming. We have therefore 
found it convenient to parameterize the $J$'s in terms of a set of 
smoothly-varying functions. Working in Euclidean space, and with the momentum 
routings of Eqs.~(\ref{qprop NLO decomp}) and (\ref{c defn}), the $J$'s 
always have a spacelike momentum argument and so are smoothly-varying 
functions. We have found that they are well represented by expansions in 
Chebyshev polynomials of the variable $x=\exp ( q^2/\Lambda^2)$. 

\begin{figure}
\begin{center}
\includegraphics[bb=0 0 567 567,width=2.5in,height=3in]{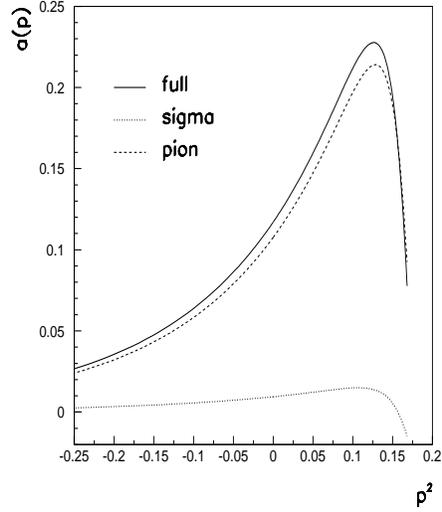}
\end{center}
\caption{The function $a(p)$ in the quark propagator at NLO, plotted against 
$p^2$ in GeV$^2$. Also shown are the contributions to $a(p)$ obtained by 
dressing the quark with pion and sigma clouds separately. The parameter set 
used is the one with $m_0(0)=300$ MeV in Table~\ref{tab1}.}
\label{q_prop_a}
\end{figure}

\begin{figure}
\begin{center}
\includegraphics[bb=0 0 567 567,width=2.5in,height=3in]{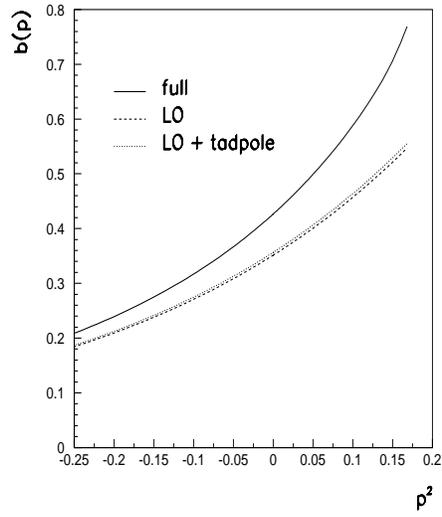}
\end{center}
\caption{The function $b(p)$ in the NLO quark propagator, plotted against 
$p^2$ in GeV$^2$. Also shown are the LO contribution and the sum of the LO 
result and the tadpole contributions. The parameter set is used is the one 
with $m_0(0)=300$ MeV in Table~\ref{tab1}.}
\label{q_prop_b}
\end{figure}

We write the inverse quark propagator as
\begin{equation}
S_F^{-1}(p)=(1+a(p)) {\rlap/p} - b(p) . \label{ab defn}
\end{equation}
The functions $a(p)$ and $b(p)$ are shown in Figs.~\ref{q_prop_a} and
\ref{q_prop_b}, for both spacelike and timelike momenta. They are plotted only
up to the energy given by the real part of the complex pole in the LO quark 
propagator. Above that energy there are pseudo-threshold effects in the model
associated with the continuation from Euclidean to Minkowski 
space\cite{nonlocal}. 

At LO $a(p)=0$ and there is no wave function renormalization. The NLO 
contributions to the coefficient ${\rlap/p}$ range up to about 0.25, which is
consistent with the expected magnitude of $1/N_c$ corrections. An interesting
aspect of the results for $a(p)$ is the appearance of a sudden dip just before
the pseudo-threshold energy. It would be interesting to examine the behaviour
of the function above this energy, although that would require a detailed 
analysis of the quark pole structure at NLO, which is beyond the scope of the
present work. Also plotted are the individual contributions to $a(p)$ which 
arise from dressing the quark line with virtual pions and with virtual sigma 
mesons. The pion cloud is obviously the dominant contributor. Since its
propagator has a pole at small timelike momentum, one expects the $T$ matrix 
in the pseudoscalar channel to be large at modest values of spacelike momenta
(the region which dominates the NLO integrals). Note also that the pion 
contributions contain an extra factor of three due to isospin.

\begin{figure}
\begin{center}
\includegraphics[bb=188 566 498 688]{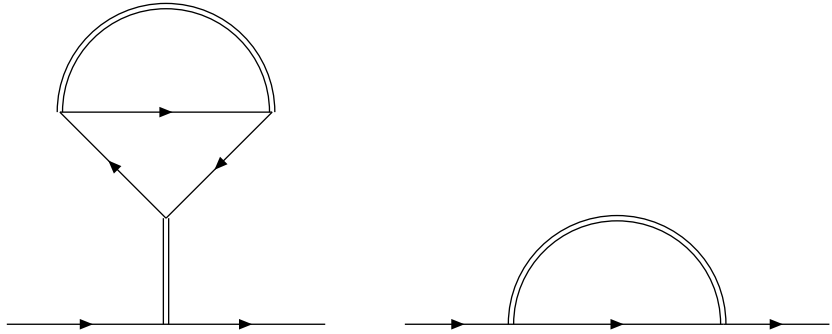}
\end{center}
\caption{Diagrams contributing to the NLO quark self-energy. Double lines are 
used to denote the LO $\overline{q}q$ scattering matrix.}
\label{NLO_q-self}
\end{figure}

The coefficient of the unit matrix in the quark self-energy at NLO receives 
contributions from both the tadpole and the meson-cloud diagrams of 
Fig.~\ref{NLO_q-self}. The results in Fig.~\ref{q_prop_b} show that the 
tadpole contribution is in fact rather small, $C \ll m(0)$. The meson-cloud 
contributions are rather more significant, increasing $b(0)$ by about $25 \%$
(again consistent with expectations for a $1/N_c$ effect). The net NLO shift 
in the quark ``mass'' at zero momentum $b(0)/(1 + a(0))$ is therefore fairly 
modest, $\sim 15 \%$. Quark dressing by meson clouds in the local NJL model 
has been investigated in Ref.\cite{QK}. It was found that the pion cloud tends
to increase $b(p)$ but that this is partially cancelled by the sigma cloud. 
The nonlocal model studied here supports that conclusion and is able to place 
it on a firmer footing since there are no ambiguities associated with the 
meson loop regularization\footnote{Note that the tadpole contributions were 
not included in Ref.\cite{QK}. Moreover the model meson propagators were
approximated by the canonical forms for point-like bosons.}.

Although the tadpole diagrams in the NLO quark self-energy are not numerically
significant, $C$ controls the NLO changes to the pion mass and decay constant.
This can be seen in the chiral expansions of these quantities described in
Appendix \ref{gmor_nlo}. One might therefore wonder whether there is some 
physical reason why its chiral limit $C_0$ turns out to be small. For the 
parameter set with $m_0(0)=200$ MeV Table~\ref{tab2} shows that the pion 
tadpole makes a contribution of about $45 \%$ of the LO chiral 
quark mass, but that this is cancelled to a large extent by the sigma tadpole.
Such cancellations do not persist at larger $m_0(0)$, where the pion tadpole 
changes sign. However for these parameter sets both tadpole contributions are 
small. In these cases it is a cancellation between ``true" virtual meson 
effects and Fock terms (similar to what happens in the quark condensate) which
is responsible for the small net effect.

In the alternative scheme of Ref.~\cite{DSTL}, the $1/N_c$ contribution of
the tadpole diagram in Fig.~\ref{NLO_q-self} is resummed to all orders when 
the SDE is solved self-consistently. In contrast, the momentum-dependent 
contribution of the meson-cloud diagram is treated perturbatively. 
This partial resummation is motivated by an expectation that contributions 
from the tadpole diagrams are larger than the other NLO effects\cite{DSTL}. 
However we find that such an expectation is not fulfilled in the nonlocal 
model; in fact we find that the contribution of the tadpole diagram is small
compared to that of the meson-cloud diagram. We thus see no compelling reason
to resum this contribution, and so we can avoid possible problems with the 
self-consistent scheme which have been noted in the context of the local NJL 
model\cite{oertel_thesis}.

In the extended version of the model there are additional contributions to the
NLO quark self-energy from tadpole and meson-cloud diagrams with other mesons.
Although a calculation of the properties of these mesons at NLO is beyond the 
scope of the present work, it is straightforward to evaluate their
contributions to the quark self-energy. As in the simpler model, we find that 
there is only a modest NLO change to the condensate and that the constant $C$ 
is small. Pseudoscalar-axial mixing makes a significant difference to the 
contribution from the pion cloud to the function $b(p)$, but this is largely 
cancelled by the vector mesons to leave a function very similar to that found 
in the simpler version of the model.

\begin{figure}
\begin{center}
\includegraphics[bb=0 0 567 567,width=2.5in,height=3in]{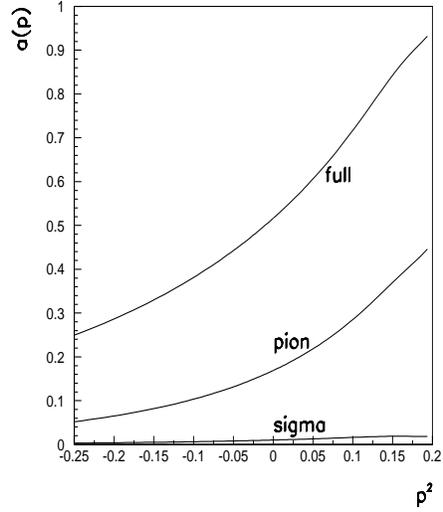}
\end{center}
\caption{The function $a(p)$ in the quark propagator of the extended model, 
plotted against $p^2$ in GeV$^2$. The full NLO result is shown together with 
the contributions from the pion and sigma clouds.}
\label{q_ext_a} 
\end{figure}

The function $a(p)$ in the extended model is shown in Fig.~\ref{q_ext_a}. 
Mixing in the pion channel and the clouds of the spin-1 states are again 
important. In $a(p)$ these effects reinforce each other and so the function 
is significantly larger than in the simpler model. An interesting difference 
between Figs.~\ref{q_prop_a} and \ref{q_ext_a} is the absence in the latter
case of a dip just before the LO pseudo-threshold energy. This is eliminated 
in the overall result due to the contributions from the longitudinal 
components of the vector and axial propagators. In particular, the steep drop 
in the pion contribution which occurs in the simpler model is removed by 
mixing with the axial channel. We note that the local NJL model has been used 
in Ref.\cite{nlo_rho} to describe the $\rho$ meson at NLO, assuming that the 
effects of $\pi a_1$ mixing and intermediate spin-1 states could be neglected.
The results presented here raise some doubts about those assumptions. Of 
course definitive statements about the quark propagator in the present model 
cannot be made without a full NLO analysis of meson properties in order to 
refit the model parameters. Since $1/N_c$ effects could alter meson masses 
significantly, the parameters used here may not constitute a reasonable choice
for the extended model at NLO.

\subsection{Pion properties}\label{num pion}

\begin{figure}
\begin{center}
\includegraphics[bb=0 0 567 567,width=2.5in,height=3in]{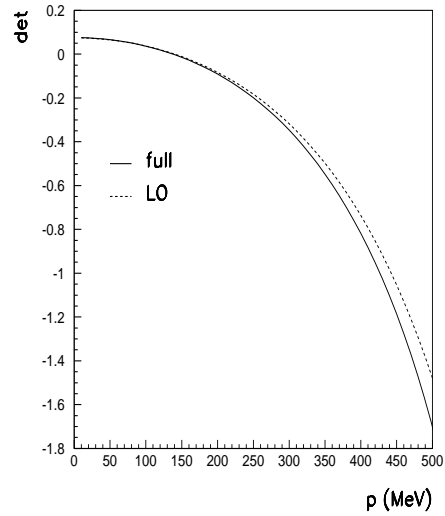}
\end{center}
\caption{The denominator of the scattering matrix in the pion channel at LO 
and NLO as a function of timelike momentum. The parameter set used is the one 
with $m_0(0)=300$ MeV in Table~\ref{tab1}.}
\label{pion_denom}
\end{figure}

The denominator of the ${\overline q}q$ scattering amplitude in the pion 
channel, Eq.~(\ref{pion shell}), is shown in Fig.~\ref{pion_denom}. For low
energies, less than about 250 MeV, the differences between the LO and NLO
curves are very small. This implies that NLO contributions to both the pion 
mass and the pion-quark coupling are extremely small. 

Although NLO contributions to the pion mass must cancel in the chiral limit, 
as described in Appendix~\ref{gmor_nlo}, such cancellations do not influence 
higher order terms in the chiral expansion. The smallness of the NLO terms in 
the pion propagator is another consequence of the cancellation between virtual
mesons and Fock terms, as we saw above in the case of the quark propagator. In
particular, all of the contributions to the pion mass and decay constant that 
are proportional to the NLO piece of the quark condensate or to $C$ are small.
(More details of these quantities are given in the appendices.) This supports 
the usual LO treatment of the pion in this type of model and it also justifies
our use of model parameters fitted to the LO pion properties in these NLO 
computations. Actual determinations of the NLO shifts in $m_\pi$ and $f_\pi$ 
have not been made in this work since they are sufficiently small (not more 
than a couple of MeV) that our numerical integration procedures would have to 
be refined in order to yield accurate values.

At higher energies, beyond those shown in Fig.~\ref{pion_denom}, NLO
contributions in the pion channel do start to become significant. When the
$\pi\sigma$ channel opens, the denominator develops an imaginary part. Shortly
before the pseudo-threshold energy (twice the real part of the complex energy 
of the pole in the quark propagator\cite{nonlocal}) is reached, the NLO terms 
become sufficiently important to change the qualitative behaviour of the pion
Bethe-Salpeter amplitude. The slope of the denominator changes sign and there
is even a second zero of Eq.~(\ref{pion shell}). This is a potentially 
worrying
feature since it corresponds to an unphysical pole in the amplitude, whose
residue has the wrong sign. A similar pole has been found in the local NJL
model\cite{oertel}. In both models the undesirable high-energy behaviour is
caused by the insertions of NLO quark self-energies on the quark lines of the
basic polarization loop integral. However it is perhaps worth noting that this
behaviour is reminiscent of that exhibited by $a(p)$ at high energies in 
Fig.~\ref{q_prop_a}. As mentioned in Sec.~\ref{num quark}, this behaviour is 
not present in the model with vector and axial mesons and it is thus possible 
that the unphysical pion pole may be removed by extending the model to 
include interactions in these channels.

\subsection{Sigma properties}\label{num sigma}

The interpretation of the scalar, isoscalar sigma meson in dynamical quark
models is complicated  by the fact that it does not correspond to an
experimentally well-determined resonance. The problem, not just for models of
such a resonance but also for phenomenological determinations of its
properties, is its very strong coupling to the two-pion channel. Indeed, while
some phenomenological analyses favour a scalar, isoscalar resonance at around 
1~GeV\cite{h_sigma}, more recently others have found resonances at 600 MeV or
lower\cite{l_sigma} (for reviews and more complete lists of references see
Refs.\cite{PDG,pennington,ML}). The strong coupling to pions means that any
resonance must form a very broad structure. Consequently the corresponding 
pole
in the scattering amplitude must lie far from the real axis, which makes its
exact position hard to determine in a model-independent way\cite{pennington}.

This strong coupling to two pions also means that a bare ${\overline q}q$ 
state in the scalar, isoscalar channel should not be directly compared with 
any phenomenological resonance. The effects of the two-pion channel need to be
included. In the present model, the leading effects of this type arise from 
the second diagram of Fig.~\ref{NLO_BSE}.

At LO in $1/N_c$, the nonlocal NJL model leads to a ${\overline q}q$ sigma 
meson with a mass of less than 500 MeV. This is similar to the local model, 
where the sigma lies on the ${\overline q}q$ threshold\cite{NJL,NJLrevs}. Some
of the properties of the state are listed in Table~\ref{tab3}. It is strongly 
coupled to two pions, although its low mass means that its decay width is less
than about 160 MeV. Although wide, this is still narrower than 
phenomenological fits which tend to give widths of at least 300 MeV, even for 
low sigma masses\cite{l_sigma,PDG,pennington,ML}.

\begin{figure}
\begin{center}
\includegraphics[bb=0 0 567 567,width=2.5in,height=3in]{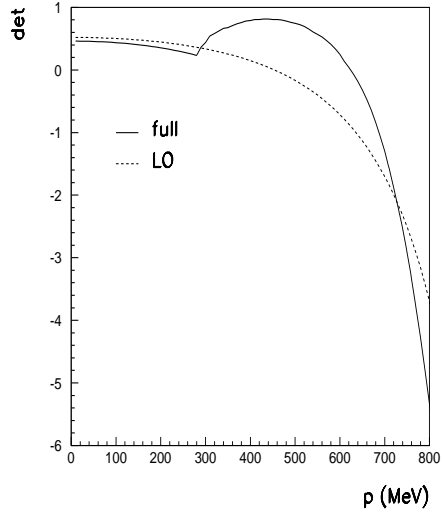}
\end{center}
\caption{The real part of the denominator of the scattering matrix in the 
sigma channel at LO and NLO as a function of timelike momentum. The parameter 
set used is the one with $m_0(0)=300$ MeV in Table~\ref{tab1}.}
\label{sigma_denom}
\end{figure}

\begin{figure}
\begin{center}
\includegraphics[bb=0 0 567 567,width=2.5in,height=3in]{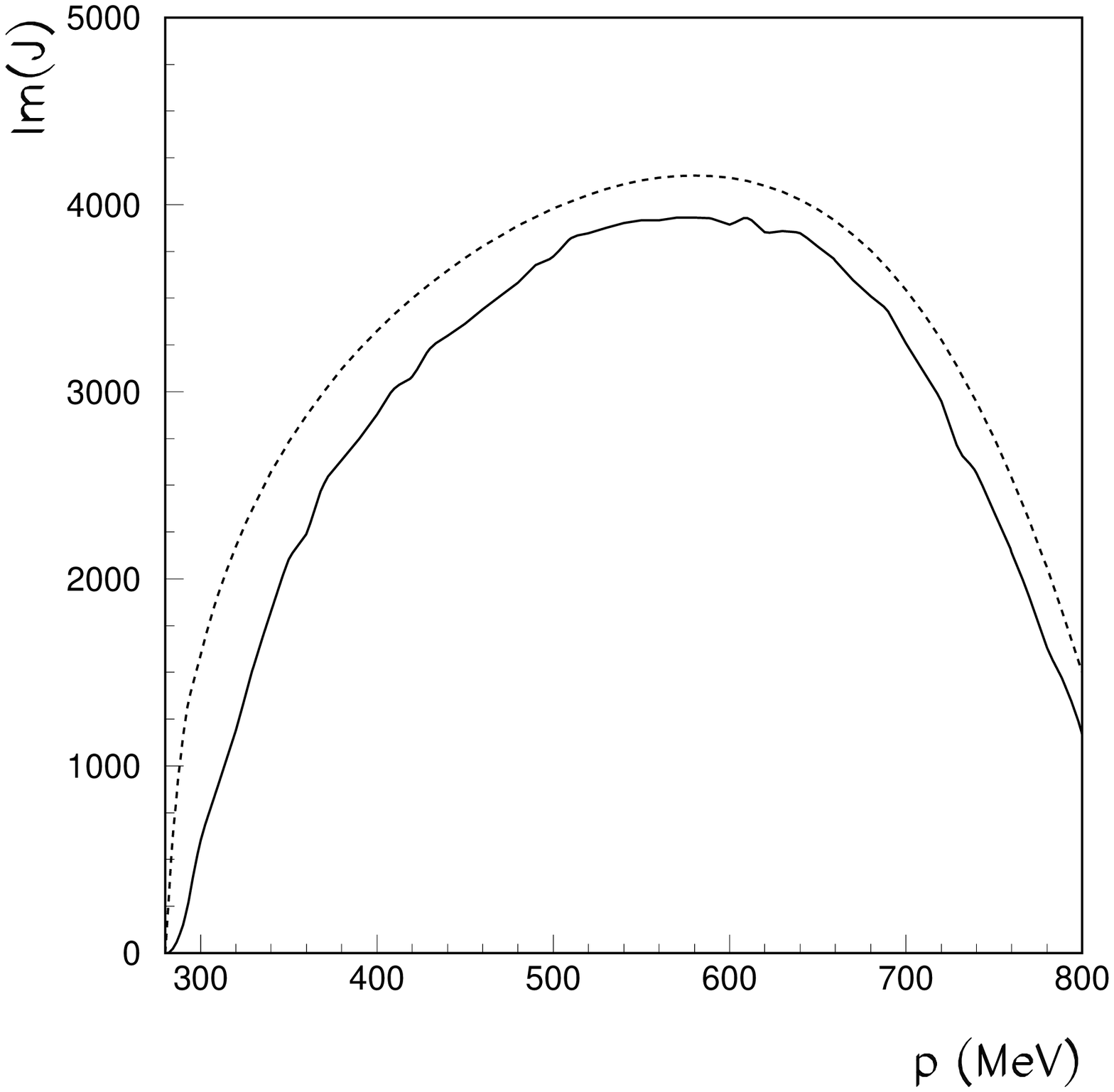}
\end{center}
\caption{The imaginary part (in MeV$^2$) at NLO of the loop $J$ in the sigma 
channel as a function of timelike momentum. The parameter set used is the one 
with $m_0(0)=300$ MeV in Table~\ref{tab1}. The solid line shows the result of
a direct evaluation; the dashed line an estimate based on the Cutkosky 
cutting rules.}
\label{sigma_imag}
\end{figure}

The real part of the denominator of the scattering amplitude at NLO is shown 
in Fig.~\ref{sigma_denom}. Above the two-pion threshold, there is an imaginary
part generated by the contribution to $J_{\sigma\sigma}$ from the second 
diagram of Fig.~\ref{NLO_BSE}. This can be estimated from the LO decay
amplitude using the Cutkosky cutting rules\cite{IZ} and it is shown in 
Fig.~\ref{sigma_imag} along with the results from a direct numerical 
evaluation. 
Some care must be taken in the evaluation of this diagram 
owing to singularities of the scattering matrices in the integrand above the 
decay threshold. We have regulated Eq.~(\ref{BSE NLO 2mes}) by replacing 
$\hat T_{\pi\pi}$ with $G_1(1-G_1J^{(0)}_{\pi\pi}-i\epsilon)^{-1}$ and 
linearly
extrapolated our results to $\epsilon=0$, based on several computations for
$\epsilon \sim 10^{-3}$.

The values listed in Table~\ref{tab3} for the mass of 
the sigma at NLO are for the mass defined as the zero of the real part of the 
denominator. Although this is a commonly used choice, it might be better to 
use the real part of the complex energy of the pole in the scattering matrix, 
as discussed by Pennington\cite{pennington}. The choice used here is purely 
one of convenience. In any case the width of the state we find is not so large
that we would expect substantial differences between the two definitions. 

We find that the sigma mass is increased from its LO value by about 30\%. 
Below the two-pion threshold the largest NLO contribution to 
$J_{\sigma\sigma}$ arises from virtual 
two-pion intermediate states. This is partially 
cancelled by the NLO quark self-energy insertions to leave a modest net
increase in $J_{\sigma\sigma}$ for most parameter sets. On their own these 
effects
of virtual mesons would tend to reduce the sigma mass. However, above the
threshold, real two-pion states make a substantial contribution with the 
opposite
sign, leading to the increase to the sigma mass. This is consistent with
the qualitative expectation that such states should be important for modelling
the sigma.

NLO corrections to the sigma mass have been calculated by Pallante\cite{P97}
using a derivative expansion of the bosonized NJL model. In that framework 
the corrections were found to be large and negative, prompting that author to 
speculate that the $1/N_c$ expansion might break down for the mass of this 
state. In the local NJL model, the large negative corrections to 
$1-G_1J_{\sigma\sigma}$ persist in calculations to all orders in momentum if
one uses Pauli-Villars regularization of the quark loops and a sharp cutoff
for the meson loop momenta\cite{oertel_thesis,priv}. The shifts in this case 
are 
sufficiently large to move the model sigma pole to negative $q^2$, producing 
an unstable vacuum. In both of these calculations, however, the results in the
scalar sector were very sensitive to the additional cut-off parameter needed 
to regularize meson loops in the local model. In contrast, the magnitude of 
the NLO mass shift which we obtain in the nonlocal model is consistent with 
expectations for a $1/N_c$ correction. Thus it would appear that the details
of the treatment of high-momentum states are important in NJL-type models.

The imaginary part of the polarization loop $J_{\sigma\sigma}$ is shown in 
Fig.~\ref{sigma_imag} and compared with the estimate obtained by applying
the Cutkosky cutting rules to graphs with two-pion intermediate states.
The difference between these curves is caused by the analytic structure of
the quark propagator, which contains additional, unphysical poles 
(Sec.~\ref{lo_review}). It is reassuring to note that these extra poles are 
located sufficiently far from the low-energy region of interest that they
do not have a large effect on the imaginary part. For other parameter
sets, with large $m_0(0)$, the extra poles lie closer to the physical
region \cite{nonlocal} and may have some influence on 
properties of the sigma meson at NLO. Since the model predictions may be
unreliable in these cases, we have preferred to quote only upper bounds
on the sigma mass for some parameter sets in Table~\ref{tab3}, limited by
the position of the first pole in the LO quark propagator.
 
For energies up to $\sim$600 MeV, the three-quark
integral in the $\sigma\pi\pi$ vertex function is only weakly dependent on the
energy. Therefore the variation of $\mathrm{Im}(J_{\sigma\sigma})$ at such 
energies is primarily a consequence the available
two-pion phase space. At higher energies the coupling
to two pions becomes much weaker\footnote{This supports a suggestion made in
Ref.\cite{nonlocal} that the weak coupling of the scalar channel to two pions 
above the sigma mass means that the broad width found for $a_1\to\sigma\pi$ in
the model need not be inconsistent with the experimental observation of a 
small width for $a_1\to\pi(\pi\pi)_s$.}. This is similar to the behaviour of 
the coupling of the sigma to two pions observed in the quark model studied in 
Ref.\cite{Efim1} (where the scalar mass was taken as a free parameter and 
chosen with the intention of interpreting the model scalar resonance as a 
heavy, narrow state). 

We note that a two-flavour NJL model of pions and sigma mesons is insufficient
for a fully realistic description of the spectrum of scalar mesons. One must 
also take account of strangeness, radial excitations, mixing with 
$\overline{q}q$ molecules or glueballs, and decays to four pion states. A 
recent attempt to include such effects can be found in Ref.\cite{celenza}.

\section{Conclusions}

We have investigated the effects of meson fluctuations on quark and meson
properties in a nonlocal NJL model\cite{BB95,nonlocal} by treating it to NLO 
in $1/N_c$. The two-body interaction between quarks has a separable form, 
similar to that suggested by instanton liquid studies\cite{DP,Musakhanov}. At
LO the model has been shown to provide successful descriptions of mesons and 
baryons\cite{nonlocal,nlsoliton}. The interaction form factors regulate both 
quark and meson loops in a natural way. As a result, no new parameters emerge 
at NLO and the contributions are unambiguous. This makes the model a 
particularly suitable one in which to study NLO effects.

The contributions to the Schwinger--Dyson and Bethe--Salpeter equations at NLO
contain the same structures as those in the local model\cite{NBCRG,DSTL}, 
apart
from the presence of interaction form factors. The NLO pieces include the 
dressing of quarks and mesons with virtual mesons and the coupling of mesons 
to two-meson states. They also include the exchange or Fock terms for the 
basic interaction between quarks. 

With a nonlocal interaction between the quarks, there are two-body terms in 
the symmetry currents\cite{BB95,nonlocal,currents} and these must be taken 
into account in the NLO contributions to the couplings of mesons to external 
currents. We have obtained the NLO pieces of the currents by first Fierz 
transforming our interaction and then applying the technique of 
Refs.\cite{BB95,nonlocal}. As a consistency check on our treatment we have 
verified that the GMOR relation holds at NLO.

In the simple version of the model with only scalar and pseudoscalar
interactions we find that the NLO contributions to a number of quantities are
small for the parameter sets of most interest (the ones with LO quark
self-energies of about 300--500 MeV). These quantities include the quark
condensate and the tadpole piece of the quark self-energy. The small size of 
the NLO pieces is a consequence of cancellations between the effects of 
virtual mesons and Fock terms, as also noted by Ripka\cite{ripka}. Similar 
cancellations mean that the NLO contributions to the pion mass and decay
constant are also small. At NLO these quantities depend not just on the NLO
pieces of the quark condensate and pion-quark coupling, but also on the
tadpole piece of the quark self-energy. The fact that this NLO tadpole energy
is small is crucial to the smallness of the corrections to pion properties. In
other cases, such as the ``meson cloud" pieces of the quark-self energy, NLO 
contributions are larger. However they typically contribute at the $\sim 25\%$
level and so give reason to hope that the $1/N_c$ expansion is a useful 
organizing principle for this model.

In this context, it is worth making a remark about the recent claim of 
Kleinert
and Van den Bossche\cite{kle} that meson fluctuation effects in the NJL model
are large enough to restore manifest chiral symmetry, at least for small 
values of $N_c$. Although our approach is based on a loop expansion, one might
have expected to see large decreases in the magnitudes of quantities like the 
quark condensate, the quark self-energy and the pion decay constant if mesonic
fluctuations were driving the system towards symmetry restoration. In fact 
pionic fluctuations do make significant contributions in this direction, but 
they are largely cancelled by the Fock terms. In the language of 
Ref.~\cite{kle}, the Fock terms contribute at NLO in $1/N_c$ to an increased
``stiffness" against mesonic fluctuations. This raises a question about the 
estimate of the critical stiffness in that work, which relies on an expression
for the stiffness obtained from the LO pion propagator. This is in addition to
the questions raised by other authors about the universality of the critical
stiffness in $3+1$ dimensions\cite{babaev} and the need to choose an 
additional
cut-off to regulate meson loops in the local NJL model\cite{oertel,ripka}.

The sigma meson is of particular interest since it can be excited by forces 
which act to restore manifest chiral symmetry, and its mass can be thought of 
as describing the forces against symmetry restoration. A light sigma seems to 
be a common feature in all NJL-type models at LO. We find that such a state is
still present when NLO effects are included, although its mass is increased by
about 30\% (as one might expect from $1/N_c$ arguments). For our preferred
parameter sets we obtain a sigma mass in range 600--650 MeV. This 
suggests that the light sigma meson is a real feature of this type of model 
and not just an artefact of LO treatments.

However, the phenomenological identification of a corresponding low-lying 
scalar, isoscalar resonance remains controversial\cite{PDG,pennington,ML}. As 
discussed above, the problem is the strong coupling of such a state to the
two-pion channel, leading to a very broad resonance. In our model we find a
width of 100--160 MeV from the LO coupling to two pions. While large, this
width is much smaller than most phenomenological 
determinations\cite{PDG,pennington,ML}. We have not yet calculated 
NLO 
contributions to the $\sigma\pi\pi$ coupling, for which terms involving
 the 
vector mesons may prove to be important. In the extended model at NLO 
there will 
be 
$t$-channel exchange of a $\rho$ meson between the pions. The attractive 
force 
between the pions generated by this mechanism is known to be important for the
sigma resonance\cite{pennington,ML} and could increase its decay width.

Finally, we have made a preliminary study of NLO effects in the extended model
with vector and axial couplings. We find that NLO contributions, in particular
from the longitudinal vector and axial channels, can improve the high-energy 
behaviour of the quark propagator. In general NLO effects seem to be 
significantly larger than in the simple version of the model, and the
convergence of the $1/N_c$ expansion may be less good as a result.

\section*{Acknowledgements}
This work was supported by the EPSRC and PPARC.

\appendix

\section{Pion decay constant at NLO}\label{fpi_nlo}

Although the NLO contributions to the pion decay constant arise from a rather
large set of diagrams, it is possible to deduce some useful cancellations
amongst them. A complete description of these cancellations would be somewhat
lengthy and so we offer only a few brief comments here. Full details can be
found in Ref.\cite{thesis}.

Obviously the diagrams that couple the local axial current to the pion all
contain an insertion of $\gamma_\mu \gamma_5$. Contracting this with $q^\mu$ 
to isolate the longitudinal component one can then substitute for ${\rlap/q} 
\gamma_5$ from the following identity:
\begin{equation}
{\rlap/q} \gamma_5 = S^{-1} (p_+) \gamma_5 + \gamma_5 S^{-1} (p_-) 
+ (m (p_+) + m (p_-)) \gamma_5 . \label{ax_id}
\end{equation}
The contributions from the first two terms here can be simplified by 
cancelling
the LO inverse propagators with quark propagators that appear in the loop
integrals. 

\begin{figure}
\begin{center}
\includegraphics[bb=163 482 503 708]{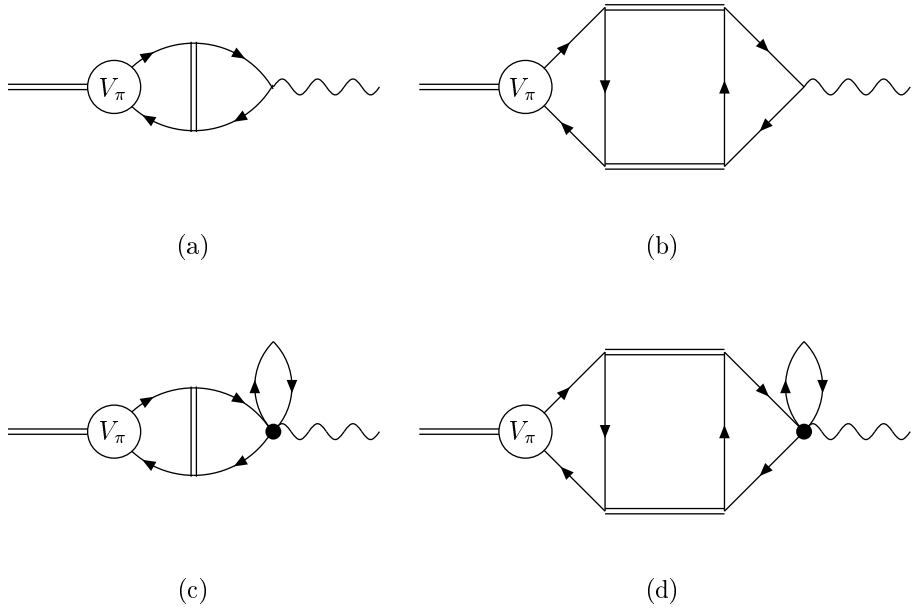}
\end{center}
\caption{NLO couplings of a pion to the axial current involving either meson 
exchange or two-meson intermediate states. There are also two-meson diagrams 
in which the direction of the arrows in one of the quark loops is reversed.}
\label{NLO_pi-ax_meson1} 
\end{figure}

Moreover, for each local diagram there is a similar nonlocal one which 
contains an additional one-quark loop. (For example, compare 
Fig.~\ref{NLO_pi-ax_meson1}a with \ref{NLO_pi-ax_meson1}c and 
Fig.~\ref{NLO_pi-ax_meson1}b with \ref{NLO_pi-ax_meson1}d.) These can be 
combined with the local contributions, as was done at LO\cite{BB95}. Each of 
the NLO nonlocal diagrams can be written as a sum of products of two-quark 
and 
one-quark loops. In one of these terms, the one-quark loop can be simplified 
by using the ladder SDE to yield a factor proportional to $m(0)-m_c$. There 
is then a cancellation between this nonlocal term and the part of the 
corresponding local diagram arising from the final term of Eq.~(\ref{ax_id}) 
which leaves only $2m_c$ surviving from the factor $m(p_+)+m(p_-)$ in 
Eq.~(\ref{ax_id}). 

\begin{figure}
\begin{center}
\includegraphics[bb=148 573 518 686]{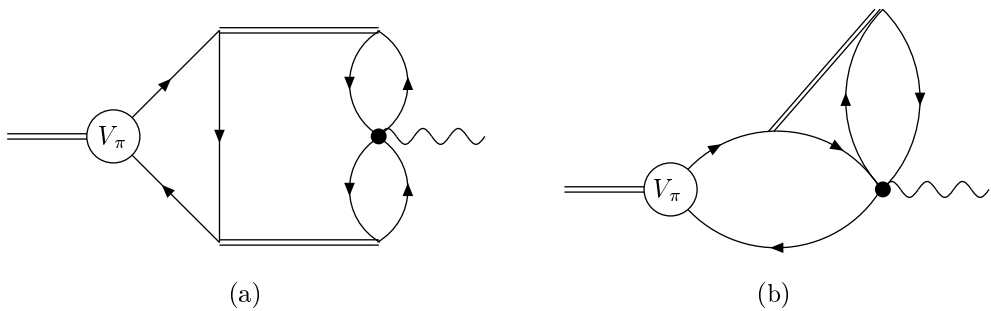}
\end{center}
\caption{NLO contributions to the coupling of a pion to the nonlocal axial 
current involving a virtual meson. There is also a second diagram similar to 
(b) in which the virtual meson couples to the antiquark line.}
\label{NLO_pi-ax_meson2} 
\end{figure}

After exploiting Eq.~(\ref{ax_id}) as outlined above, several other useful 
cancellations can be identified. As an example of the procedure, consider the 
diagram shown in Fig.~\ref{NLO_pi-ax_meson1}b. When Eq.~(\ref{ax_id}) is 
substituted for ${\rlap/q} \gamma_5$ in the triangular loop integral, the 
first two 
terms reduce to a sum of two-quark loops. Up to a factor of $-GJ$ the
resulting contributions are then similar in form to those of the diagram 
shown in Fig.~\ref{NLO_pi-ax_meson2}a. 

Hence, in the sum of the contributions from Fig.~\ref{NLO_pi-ax_meson1}b and
\ref{NLO_pi-ax_meson2}a, one can cancel one of the intermediate meson 
propagators to leave a product of a two-quark loop, a three-quark loop and one
meson propagator. It turns out that this can then be profitably combined with 
the contribution of Fig.~\ref{NLO_pi-ax_meson2}b, which has the same 
structure.

We give here the result of a full analysis of the NLO contributions to $f_\pi$
in the version of the model with only scalar and pseudoscalar interactions. 
Note that, rather than defining further loop integrals whose structure is very
similar to those given above in Sec.~\ref{BSE_NLO}, we simply indicate how the
products of form factors need to be changed in those integrals. The final 
expression for $f_\pi$ at NLO is\cite{thesis}
\begin{eqnarray}
f_{\pi}^{(1)} (q^2)& =& \frac{i g_{\pi qq}^{(0)} G_1}{2 q^2} 
J_{\pi\pi}^{(1)} (q^2) \int \frac{d^4k}{(2\pi)^4} {\mathrm{Tr}} \left[ S(k) 
\right] f(k) \Bigl( f(k+q) + f(k-q) \Bigr)\nonumber\\
&&+ \frac{g_{\pi qq}^{(0)}m_c}{q^2} \Bigl( J_{\pi\pi}^{(1)} \; 
\hbox{from Eq.~(\ref{BSE NLO qprop}) with} \; f^2(p_+) f^2(p_-) 
\to f(p_+) f(p_-) \Bigr)\nonumber\\
&&+ \frac{g_{\pi qq}^{(0)}m_c}{q^2} \Bigl( J_{\pi\pi}^{(1)} \; 
\hbox{from Eq.~(\ref{BSE NLO 1mes}) with} \; f^2(p_+) f^2(p_-) 
\to f(p_+) f(p_-) \Bigr)\nonumber\\
&&+ \frac{g_{\pi qq}^{(0)}m_c}{q^2} \Bigl( J_{\pi\pi}^{(1)} \; 
\hbox{from Eq.~(\ref{BSE NLO 2mes}) with} \; f^2(p_+) f^2(p_-) 
\to f(p_+) f(p_-) \nonumber\\
&&\qquad\qquad\qquad \hbox{in} \; \overline{L}_{\pi,km} \Bigr)\nonumber\\
&&- i \frac{C g_{\pi qq}^{(0)}}{2q^2} \Bigl(1 - G_1 J_{\pi\pi}^{(0)} (q^2)
\Bigr) \int \frac{d^4p}{(2\pi)^4} {\mathrm{Tr}} \left[ S(p) S(p) \right] 
f^3(p) \nonumber\\
&&\qquad\qquad\qquad\qquad\qquad\qquad\qquad\qquad \times 
\Bigl( f(p+q) + f(p-q) \Bigr)\nonumber\\
&&-\frac{g_{\pi qq}^{(0)}}{2q^2} \Bigl(1 - G_1 J_{\pi\pi}^{(0)} (q^2)\Bigr) 
\sum_{i,j} \int \frac{d^4p}{(2\pi)^4} \frac{d^4k}{(2\pi)^4} \hat{T}_{ij} (p-k)
\nonumber\\ 
&&\qquad\qquad\qquad\qquad\qquad\qquad \times {\mathrm{Tr}} \left[ S(p) 
\Gamma_i S(k) \overline{\Gamma}_j S(p) \right] \nonumber\\
&&\qquad\qquad\qquad\qquad\qquad\qquad \times f^3(p) \Bigl( f(p+q) + f(p-q) 
\Bigr)f^2(k)\nonumber\\
&&+\frac{g_{\pi qq}^{(1)}}{g_{\pi qq}^{(0)}} \, f_\pi^{(0)}(q^2)  . 
\label{final fpi}
\end{eqnarray}

\section{Chiral relations}\label{gmor_nlo}

A large number of diagrams are required in a NLO treatment of the nonlocal 
NJL 
model, as described in Sec.~\ref{NLO diagms}. It is important to have some
check that a consistent set has been identified and evaluated. To this end, 
we 
demonstrate that low-energy chiral constraints are satisfied in the model at 
NLO. (Again we consider only the simpler version of the model, without vector 
and axial interactions.)

First consider the pion mass. The chiral expansion of the pion denominator,
Eq.~(\ref{pion shell}), at LO is\cite{BB95}
\begin{equation}
1 - G_1 J_{\pi\pi}^{(0)} (q^2) = - G_1 m_c \frac{\langle \overline{\psi} \psi
\rangle_0^{(0)}}{m_0(0)^2} - G_1 \frac{q^2}{g_{\pi qq0}^{(0)\,2}} + {\cal O} 
(q^4, m_c^2), \label{J_PP expansion}
\end{equation}
where the subscript $0$ is used to denote a quantity evaluated in the chiral
limit. (We apologize for the fact that the notation inevitably becomes rather 
cumbersome at this point.) If one substitutes the LO chiral expansion into the
NLO on-shell condition, it is immediately clear that the chiral expansion of 
$J_{\pi\pi}^{(1)}(q^2)$ must start at ${\cal O} (q^2, m_c)$ in order for the 
pion to remain a Goldstone boson in the chiral limit. 

A proof of this statement in the local NJL model was presented in 
Ref.\cite{DSTL}. Although one has to keep track of interaction form factors,
the proof for the nonlocal model proceeds along very similar 
lines\cite{thesis}. In particular, when one takes the chiral limit of the 
contributions to $J_{\pi\pi}^{(1)}$ from Eqs.~(\ref{BSE NLO qprop}) and
(\ref{BSE NLO 2mes}), each of the Dirac traces yields a factor that cancels 
with the denominator of one of the LO quark propagators. The three-quark loop 
in the contribution of the tadpole term to Eq.~(\ref{BSE NLO qprop}) can be
reduced to a two-quark integral which can then be used to cancel the 
denominator of the $\sigma (0)$ propagator from Eq.~(\ref{c defn}). In the 
case
of the contribution to $J_{\pi\pi}^{(1)}$ from Eq.~(\ref{BSE NLO 1mes}), the 
Dirac trace takes an even more convenient form in the chiral limit. For both 
the sigma and pion exchanges, it factorizes into pieces which cancel the 
denominators from two LO quark propagators.

An important property noted in the proof of Ref.\cite{DSTL} concerns the 
second diagram of Fig.~\ref{NLO_BSE} which involves a virtual pion and a 
virtual sigma meson. In the chiral limit the three-quark loop integrals are 
proportional to the difference between LO polarization loops: 
$J_{\sigma\sigma}^{(0)} - J_{\pi\pi}^{(0)}$. This allows one to replace the 
product of the virtual meson propagators with their difference. It is then 
possible to combine the contribution from this diagram with other 
contributions, which contain only a single meson propagator. The same 
property 
holds in the nonlocal model\footnote{It is in fact straightforward to
generalize the proof that $J_{\pi\pi0}^{(1)}(0)=0$ to the extended version of
the model, including other mesons. In so doing the products of meson
propagators occurring in diagrams of the type shown on the right--hand side of
Fig.~\ref{NLO_BSE} can be dealt with using the relations 
$J_{\sigma\sigma}^{(0)}(q^2) - J_{\pi\pi}^{(0)}(q^2) = J_{VV}^{(0)T}(q^2) - 
J_{AA}^{(0)T}(q^2) = J_{VV}^{(0)L}(q^2) - J_{AA}^{(0)L}(q^2)$, where $T$ and 
$L$ denote transverse and longitudinal components respectively. Note, however,
that in the extended model, one must take $\pi a_1$ mixing into account and so
to complete the proof that the pion is a Goldstone boson at NLO, it is also 
necessary to establish that $J_{A\pi 0}^{(1)}(0)=0$.}.

Many of the simplifications that lead to the vanishing of $J_{\pi\pi}^{(1)}$ 
in the chiral limit can also be exploited in calculating its ${\cal O} (m_c)$ 
and ${\cal O} (q^2)$ terms. One finds that\cite{thesis}
\begin{equation}
J_{\pi\pi}^{(1)} = -m_c 
\frac{\langle\overline{\psi}\psi\rangle_0^{(1)}}{m_0^2(0)} - m_c 
\frac{\langle \overline{\psi} \psi \rangle_0^{(0)}}{m_0^2(0)}\, 
\frac{2C_0}{m_0(0)} - q^2\, \frac{2g_{\pi qq 0}^{(1)}}{g_{\pi qq 0}^{(0)\,3}} 
+ {\cal O} (q^4, m_c^2) \label{JNPP expansion}
\end{equation}
where the NLO contribution to the condensate is given by:
\begin{equation}
\langle \overline{\psi} \psi \rangle^{(1)} = -i {\mathrm{Tr}} \int 
\frac{d^4p}{(2\pi)^4} S(p) \Sigma^{(1)} (p) S(p) .
\end{equation} 
Substituting the above expansion into the on-shell condition 
(Eq.~(\ref{pion shell})) yields the pion mass to this order,
\begin{equation}
m_\pi^2= -\left(g_{\pi qq 0}^{(0)} + g_{\pi qq 0}^{(1)}\right)^2 
\frac{m_c[ \langle \overline{\psi} \psi \rangle_0^{(0)} 
+ \langle \overline{\psi} \psi \rangle_0^{(1)} ]}{(m_0(0) + C_0)^2} 
+ {\cal O} (m_c^2) + {\cal O} (N_c^{-2}). \label{mpi NLO shift}
\end{equation}

We see that the chiral expansion of the pion Bethe--Salpeter amplitude at NLO
retains a similar structure to that at LO. The changes include the expected 
ones from the NLO contributions to the quark condensate and the pion-quark 
coupling. However, the form of the dynamical mass scale that appears in 
Eq.~(\ref{mpi NLO shift}) is less obvious, the shift in this being given 
entirely by the contribution of the tadpole diagram, Eq.~(\ref{c defn}). This 
is despite the fact that the meson cloud of a dressed quark contributes a 
term 
of ${\cal O} (m_c)$ to the chiral expansion of $J_{\pi\pi}^{(1)}$ and also 
makes a 
finite contribution to the quark self-energy at zero momentum. The fact that 
the NLO shift in the pion mass is controlled by the coefficient of the 
tadpole 
diagram has also been noted in the local NJL model\cite{NBCRG,nlo_rho}, in 
which the overall mass scale can be recognized as the stationary point of the 
one-meson-loop effective action\cite{NBCRG,newpreprint}.

From Eq.~(\ref{mpi NLO shift}) it can be seen that the 
Gell-Mann--Oakes--Renner
relation\cite{GMOR} will be satisfied if a version of the Goldberger--Treiman
relation\cite{GT} holds in the chiral limit at NLO,
\begin{eqnarray}
f_{\pi 0}^{(0)} + f_{\pi 0}^{(1)} &=& \frac{m_0(0) + C_0}{g_{\pi qq 0}^{(0)} 
+ g_{\pi qq 0}^{(1)}} \nonumber\\
&=&\frac{m_0(0)}{g_{\pi qq 0}^{(0)}} 
- \frac{g_{\pi qq 0}^{(1)}}{g_{\pi qq 0}^{(0)}}\, 
\frac{m_0(0)}{g_{\pi qq 0}^{(0)}} + \frac{C_0}{g_{\pi qq 0}^{(0)}} 
+ {\cal O} (N_c^{-2}). \label{gt NLO}
\end{eqnarray}
The term of order $N_c^0$ on the right-hand side of this condition was shown 
in Ref.\cite{BB95} to be equal to the LO part of $f_\pi$. 

In order to demonstrate that the NLO part of the decay constant reduces to 
the 
next two terms of Eq.~(\ref{gt NLO}) in the chiral limit, first note that all 
of the contributions in Eq.~(\ref{final fpi}) are of zeroth order in the 
chiral expansion. Hence in all of the integrals the chiral limit may be taken 
directly, without making an expansion of the integrand. In the first term of 
Eq.~(\ref{final fpi}) the ladder SDE can be used to perform the integral, 
giving
\begin{equation}
-2 m_0(0) \frac{g_{\pi qq 0}^{(1)}}{g_{\pi qq 0}^{(0)\,2}} 
+ m_c \frac{g_{\pi qq 0}^{(0)}}{q^2}\, 
\frac{\langle \overline{\psi} \psi \rangle_0^{(1)}}{m_0(0)} 
- m_c \frac{g_{\pi qq 0}^{(0)}}{q^2}\, 
\frac{\langle \overline{\psi} \psi \rangle_0^{(0)}}{m_0(0)}\, 
\frac{2C_0}{m_0(0)}  \label{J NLO}
\end{equation}
where Eq.~(\ref{JNPP expansion}) has been used for the chiral expansion of 
$J_{\pi\pi}^{(1)}$. 

The next three terms in Eq.~(\ref{final fpi}) are explicitly proportional to
$m_c$ and contain integrals which are very similar to those occurring in 
$J_{\pi\pi}^{(1)}$, but with different combinations of form factors. 
Procedures for manipulating such integrals in order to simplify their sum were
described in Ref.\cite{DSTL} and also above, where they were exploited in 
order to show that $J_{\pi\pi}^{(1)}$ vanishes in the chiral limit. Very 
similar simplifications can be made here, despite the different form factor 
structures that appear. These give rise to a nonvanishing contribution to 
$f_{\pi0}^{(1)}$, 
\begin{equation}
\frac{g_{\pi qq 0}^{(0)}m_c}{q^2m_0(0)} 
\left( - \langle \overline{\psi} \psi \rangle_0^{(1)} 
+ \frac{C_0}{m_0(0)} \langle \overline{\psi} \psi \rangle_0^{(0)} \right) \, .
\end{equation}

The fifth term in Eq.~(\ref{final fpi}) is proportional to the tadpole
contribution $C$. Using Eq.~(\ref{J_PP expansion}) to expand the factor of 
$1-G_1 J_{\pi\pi}^{(0)}(q^2)$, the chiral limit of this term is 
\begin{equation}
\frac{g_{\pi qq 0}^{(0)}}{q^2} C_0 G_1 J_{\sigma\sigma 0}^{(0)}(0) 
\left( \frac{m_c \langle \overline{\psi} \psi \rangle_0^{(0)}}{m_0^2(0)} 
+ \frac{q^2}{g_{\pi qq 0}^{(0)\,2}} \right) . \label{fpi chi 1}
\end{equation}

In the chiral limit, the integral in the sixth term of Eq.~(\ref{final fpi}) 
reduces to the one in the definition of $C$, Eq.~(\ref{c defn}), giving
\begin{equation}
\frac{g_{\pi qq 0}^{(0)}}{q^2} C_0 \Bigl(1 - G_1 J_{\sigma\sigma 0}^{(0)}(0)
\Bigr) 
\left( \frac{m_c \langle \overline{\psi} \psi \rangle_0^{(0)}}{m_0^2(0)} 
+ \frac{q^2}{g_{\pi qq 0}^{(0)\,2}} \right) . \label{fpi chi 11}
\end{equation}
The final term of Eq.~(\ref{final fpi}) is trivially 
\begin{equation}
\frac{g_{\pi qq 0}^{(1)}}{g_{\pi qq 0}^{(0)}}\, 
\frac{m_0(0)}{g_{\pi qq 0}^{(0)}} . \label{g NLO}
\end{equation}

Adding together Eqs.~(\ref{J NLO}) to~(\ref{g NLO}) reproduces exactly the 
${\cal O} (N_c^{-1})$ terms of Eq.~(\ref{gt NLO}), demonstrating that the
Gell-Mann--Oakes--Renner relation holds in our treatment of the nonlocal NJL
model at NLO.

\newpage
\section*{Tables}

\begin{table}
\begin{center}
\begin{tabular}{cdddccc}
$m_0(0)$\ &\ $G_1$(GeV$^{-2}$)\ &\ $m_c$\ &\ $\Lambda$\ &\ $m(0)$\ &
\ $-\langle\bar qq\rangle^{1/3}$\ &\ $g_{\pi qq}$\ \\
\hline
200&14.3&4.8&1459&245&246&2.56\\
250&30.5&7.8&1064&298&210&3.13\\
300&53.8&11.0&861&351&189&3.70\\
350&85.9&14.2&734&406&173&4.28\\
400&128.1&17.5&647&461&162&4.87\\
450&181.7&20.8&583&516&153&5.46\\
500&248.0&24.1&535&572&146&6.04\\
\end{tabular}
\end{center}
\caption{Values of the model parameters, fitted at LO. Also listed are the 
dynamical quark mass, the quark condensate in the chiral limit and the 
pion-quark coupling. Apart from $G_1$ and $g_{\pi qq}$, which is 
dimensionless, all quantities are given in MeV.}
\label{tab1}
\end{table}

\begin{table}
\begin{center}
\begin{tabular}{ccdddd}
$m_0(0)$\ &\ $-\langle\bar qq\rangle^{1/3}$\ &\ $C$\ &\ $C_0$\ &
\ $C_0 (\pi)$\ &\ $C_0 (\sigma)$\\
\hline
200&259&25.0&25.3&90.2&$-$64.9\\
250&215&12.4&8.0&44.2&$-$36.2\\
300&190&5.2&$-$1.7&22.4&$-$24.1\\
350&174&0.005&$-$8.8&8.6&$-$17.4\\
400&162&$-$4.1&$-$14.6&$-$1.4&$-$13.1\\
450&153&$-$7.7&$-$19.6&$-$9.5&$-$10.2\\
500&145&$-$10.8&$-$24.2&$-$16.3&$-$7.9\\
\end{tabular}
\end{center}
\caption{Properties of the quark propagator at NLO. All quantities are quoted 
in MeV. The quark condensate is evaluated in the chiral limit. The constant 
$C$ appears in the part of the self-energy from the NLO tadpole diagram and is
defined in Eq.~(\ref{c defn}). Its value in the chiral limit is $C_0$. The 
contribution to $C$ from the tadpole diagram with a virtual meson $m$ is 
denoted $C(m)$.}
\label{tab2}
\end{table}

\begin{table}
\begin{center}
\begin{tabular}{cccdc}
$m_0(0)$\ &\ $m_\sigma^{(0)}$\ &\ $g_{\sigma\pi\pi}^{(0)}$\ &
\ $\Gamma^{(0)} (\sigma \to \pi\pi)$\ &\ Re$[m_\sigma]$\ \\
\hline
200&385&1092&63.5&455\\
250&423&1336&94.4&552\\
300&454&1562&126.3&624\\
350&477&1732&152.0&650\\
400&492&1783&158.7&$>$574\\
450&489&1489&111.0&$>$504\\
500&478&1515&116.2&$>$450\\
\end{tabular}
\end{center}
\caption{Sigma meson mass evaluated at NLO. For comparison the mass and other 
properties at LO are also given (denoted by a superscript $(0)$). All 
quantities are quoted in MeV. For the parameter sets with large $m_0(0)$, the 
energy at which pseudo-threshold effects are 
required\protect\cite{nonlocal} is rather
modest, and hence we quote only a lower bound for the mass at NLO.}
\label{tab3}
\end{table}

\newpage
\section*{Figures}

\clearpage

\begin{thebibliography}{00}
\bibitem{NJL}Y. Nambu and G. Jona-Lasinio, Phys.\ Rev.\ {\bf 122} (1961) 345;
{\bf 124} (1961) 246.
\bibitem{NJLrevs}U. Vogl and W. Weise, Prog.\ Part.\ Nucl.\ Phys.\ {\bf 27}
(1991) 195; S. P. Klevansky, Rev.\ Mod.\ Phys.\ {\bf 64} (1992) 649; T. 
Hatsuda
and T. Kunihiro, Phys.\ Rep.\ {\bf 247} (1994) 221; J. Bijnens, Phys.\ Rep.\
{\bf 265} (1996) 369.
\bibitem{QK}E. Quack and S. P. Klevansky, Phys.\ Rev.\ {\bf C49} (1994) 3283.
\bibitem{nlo_rhodecay}A. Polleri, R. A. Broglia, P. M. Pizzochero and N. N.
Scoccola, Z.\ Phys.\ {\bf A357} (1997) 325.
\bibitem{nlo}N.-W. Cao, C. M. Shakin and W.-D. Sun, Phys.\ Rev.\ {\bf C46}
(1992) 2535; P. P. Domitrovich, D. B\"{u}ckers and H. M\"{u}ther, Phys.\ Rev.\
{\bf C48} (1993) 413; P. Zhuang, J. H\"{u}fner and S. P. Klevansky, Nucl.\
Phys.\ {\bf A576} (1994) 525; P. Zhuang, J. H\"{u}fner, S. P. Klevansky and H.
Voss, Ann.\ Phys.\ (N.Y.) {\bf 234} (1994) 225; Y. B. He, J. H\"{u}fner, S. P.
Klevansky and P. Rehberg, Nucl.\ Phys.\ {\bf A630} (1998) 719.
\bibitem{NBCRG}E. Nikolov, W. Broniowski, C. V. Christov, G. Ripka and K.
Goeke, Nucl.\ Phys.\ {\bf A608} (1996) 411.
\bibitem{Bowler}R. D. Bowler, Ph.D.\ thesis, University of Manchester, 1995.
\bibitem{DSTL}V. Dmitra\u{s}inovi\'{c}, H.-J. Schulze, R. Tegen and R. H.
Lemmer, Ann.\ Phys.\ (N.Y.) {\bf 238} (1995) 332.
\bibitem{P97}E. Pallante, Z.\ Phys.\ {\bf C75} (1997) 305.
\bibitem{nlo_chi}M. Huang, P. Zhuang and W. Chao, Commun.\ Theor.\ Phys.\ 
{\bf 34} (2000) 91.
\bibitem{qmatter}B. Szczerbinska and W. Broniowski, Acta Phys.\ Polon.\ 
{\bf B31} (2000) 835; D. G\'{o}mez Dumm and N. N. Scoccola, hep-ph/0107251 
(2001).
\bibitem{oertel}M. Oertel, M. Buballa and J. Wambach, Phys.\ Lett.\ {\bf B477}
(2000) 77.
\bibitem{nlo_rho}M. Oertel, M. Buballa and J. Wambach, Nucl.\ Phys.\ 
{\bf A676} (2000) 247.
\bibitem{ripka}G. Ripka, Nucl.\ Phys.\ {\bf A683} (2001) 463; hep-ph/0007250 
(2000).
\bibitem{oertel_thesis}M. Oertel, Ph.D.\ thesis, University of Darmstadt, 
2000.
\bibitem{kle}H. Kleinert and B. Van den Bossche, Phys.\ Lett.\ {\bf B474} 
(2000) 336; hep-ph/9908284 (1999).
\bibitem{newpreprint}M. Oertel, M. Buballa and J. Wambach, Phys.\ Atom.\ 
Nucl.\ {\bf 64} (2001) 698; Yad.\ Fiz.\ {\bf 64} (2001) 757.
\bibitem{BB95}R. D. Bowler and M. C. Birse, Nucl.\ Phys.\ {\bf A582} (1995) 
655.
\bibitem{nonlocal}R. S. Plant and M. C. Birse, Nucl.\ Phys.\ {\bf A628} (1998)
607.
\bibitem{Broniowski}W. Broniowski, AIP Conf.\ Proc.\ {\bf 508} (2000) 380.
\bibitem{nlsoliton}B. Golli, W. Broniowski and G. Ripka, Phys.\ Lett.\ 
{\bf B437} (1998) 24; W. Broniowski, hep-ph/9909438 (1999); G. Ripka and B. 
Golli, AIP Conf.\ Proc.\ {\bf 508} (2000) 3; W. Broniowski, B. Golli and G. 
Ripka, hep-ph/0107139 (2001).
\bibitem{DP}D. I. Dyakonov and V. Yu.\ Petrov, Nucl.\ Phys.\ {\bf B245} (1984)
259; Sov.\ Phys.\ JETP {\bf 62} (1985) 204, 431; Nucl.\ Phys.\ {\bf B272}
(1986) 457.   
\bibitem{Musakhanov}M. Musakhanov, Eur.\ Phys.\ J. {\bf C9} (1999) 235.
\bibitem{currents}K. Naito, K. Yoshida, Y. Nemoto, M. Oka and M. Takizawa,
Phys.\ Rev.\ {\bf C59} (1999) 1095.
\bibitem{GMOR}M. Gell-Mann, R. Oakes and B. Renner, Phys.\ Rev.\ {\bf 175}
(1968) 2195.
\bibitem{thesis}R. S. Plant, Ph.D.\ thesis, University of Manchester, 1998.
\bibitem{schmidt}D. Blaschke, Yu. L. Kalinovsky, G. R\"{o}pke, S. Schmidt and 
M. K. Volkov, Phys.\ Rev.\ {\bf C53} (1996) 2394.
\bibitem{IZ}C. Itzykson and J.-B. Zuber, {\em Quantum Field Theory\/}
(McGraw-Hill 1980).
\bibitem{h_sigma}K. L. Au, D. Morgan and M. Pennington, Phys.\ Rev.\ {\bf D35}
(1987) 1633; D. Morgan and M. R. Pennington, Phys.\ Rev.\ {\bf D48} (1993) 
1185.
\bibitem{l_sigma}B. S. Zou and D. V. Bugg, Phys.\ Rev.\ {\bf D50} (1994) 591;
R. Kaminski, L. Lesniak and J.-P. Maillet, Phys.\ Rev.\ {\bf D50} (1994) 
3145; 
S. Ishida {\em et al.}, Prog.\ Theor.\ Phys.\ {\bf 95} (1996) 745; {\em ibid.}
{\bf 98} (1997) 1005; N. A. Tornqvist and M. Roos, Phys.\ Rev.\ Lett.\ 
{\bf 76} (1996) 1575; M. Harada, F. Sannino and J. Schechter, Phys.\ Rev.\ 
{\bf D54} (1996) 1991; R. Kaminski, L. Lesniak and B. Loiseau, Phys.\ Lett.\ 
{\bf B413} (1997) 130; Eur.\ Phys.\ J. {\bf C9} (1999) 141; T. Hannah, Phys.\ 
Rev.\ {\bf D60} (1999) 017502; M. Y. Ishida {\em et al.}, hep-ph/9905261 
(1999); V.V. Anisovich and V. A. Nikonov, hep-ph/9911512 (1999).
\bibitem{PDG}Particle Data Group, Eur.\ Phys.\ J. {\bf C3} (1998) 1.
\bibitem{pennington}M. R. Pennington, hep-ph/9905241 (1999).
\bibitem{ML}V. E. Markushin and M. P. Locher, hep-ph/9906249 (1999).
\bibitem{priv}M. Oertel (private communication).
\bibitem{Efim1}G. V. Efimov and M. A. Ivanov, {\em The Quark Confinement Model
of Hadrons\/} (IOP, Bristol and Philadelphia, 1993).
\bibitem{celenza}L. S. Celenza, B. Huang, H. Weng and C. M. Shakin, Phys.\ 
Rev.\
{\bf C60} (1999) 065210; L. S. Celenza, S.-F. Gao, B. Huang, H. Weng and C. M.
Shakin, Phys.\ Rev.\ {\bf C61} (2000) 035201. 
\bibitem{babaev}E. Babaev, Phys.\ Rev.\ {\bf D62} (2000) 074020.
\bibitem{GT}M. Goldberger and S. Treiman, Phys.\ Rev.\ {\bf 110} (1958) 1178,
1478.

\end{thebibliography}
\end{document}